\documentclass[%
 aip,
 jcp,%
amsmath,amssymb,
reprint,%
noshowkeys,
]{revtex4-1}

\usepackage{graphicx}
\usepackage{dcolumn}
\usepackage{bm}
\usepackage{epstopdf}
\usepackage{color}
\usepackage{ulem}
\usepackage{amsmath}
\usepackage{amsthm}

\newcommand*{\jk}{{j_1,\dots,j_K}}
\newcommand*{\jkp}{{j_1,\dots,j_k+1,\dots,j_K}}
\newcommand*{\jkm}{{j_1,\dots,j_k-1,\dots,j_K}}
\usepackage{hyperref}

\newtheoremstyle{query}%
{}{}
{\color{red}}
{}
{\sffamily\bfseries}{:}{12pt}
{}

\begin{document}

\title{Perspective: Numerically ``exact'' approach to open quantum dynamics: The hierarchical equations of motion (HEOM)}
\thanks{\href{https://doi.org/10.1063/5.0011599}{https://doi.org/10.1063/5.0011599}}

\author{Yoshitaka Tanimura}
\email{tanimura.yoshitaka.5w@kyoto-u.jp}
\affiliation{Department of Chemistry, Graduate School of Science, Kyoto University, Kyoto 606-8502, Japan}
\date{\today} 

\begin{abstract}
An open quantum system refers to a system that is further coupled to a bath system consisting of surrounding radiation fields, atoms, molecules, or proteins. The bath system is typically modeled by an infinite number of harmonic oscillators. This system--bath model can describe the time-irreversible dynamics through which the system evolves toward a thermal equilibrium state at finite temperature. In nuclear magnetic resonance and atomic spectroscopy, dynamics can be studied easily by using simple quantum master equations under the assumption that the system--bath interaction is weak (perturbative approximation) and the bath fluctuations are very fast (Markovian approximation). However, such approximations cannot be applied in chemical physics and biochemical physics problems, where environmental materials are complex and strongly coupled with environments. The hierarchical equations of motion (HEOM) can describe numerically ``exact'' dynamics of a reduced system under nonperturbative and non-Markovian system--bath interactions, which has been verified on the basis of  exact analytical solutions (non-Markovian tests) with any desired numerical accuracy.  The HEOM theory has been used to treat  systems of practical interest, in particular to account for various  linear and nonlinear spectra in molecular and solid state materials, to evaluate  charge and exciton transfer rates in biological systems, to simulate resonant tunneling and quantum ratchet processes in nanodevices, and to explore  quantum entanglement states in quantum information theories. This article, presents an overview of the HEOM theory, focusing on its theoretical background and applications, to help further the development of the study of open quantum dynamics. 
\end{abstract}

                         

\maketitle

\section{INTRODUCTION}
Time irreversibility is not a problem to be solved, but a reality to be experienced. 
This is true for the physical, chemical, and biological phenomena that we encounter throughout  our lives.
Theoretically, in particular in the quantum case,  realization of time irreversibility is difficult, because the fundamental kinetic equations,   including the Schr\"odinger equation and the Dirac equation, ensure that the dynamics are reversible in time. ``Open quantum dynamics''  refers to the dynamics of a system that is coupled to a bath system, for example, a surrounding radiation field or atomic or molecular environment. The bath system is typically modeled by an infinite number of harmonic oscillators. This system--bath model can describe the time irreversibility of the dynamics toward the thermal equilibrium state, in which the energy supplied by  fluctuations and the energy lost through dissipation are balanced, while the bath temperature does not change, because its heat capacity is infinite. 

Historically, studies of open quantum dynamics were motivated by practical consideration, such as   line-shape analysis in nuclear magnetic resonance (NMR) \cite{Bloch53,Redfield65, Kubo69} and maser and laser spectra,\cite{Lamb67,Mollow69} or by philosophical interest in  thermodynamic behavior in a quantum regime as an aspect of nonequilibrium statistical physics.\cite{Zwanzig64, Haake73} In the former context, theories have been developed to construct models to describe chemical reactions,\cite{Wolyness1981,Miller1989, HanggiRMP90} electron and charge transfer rates,\cite{MarcusRMP90, Khun95, YangTonuOliverRev2015} nonadiabatic transitions,\cite{Garg86, Sparpagilione1988} quantum device systems,\cite{MasonHess89}  ratchet rectification,\cite{Hanggi97,Hanggi09}  and superconducting quantum interference device (SQUID)rings,\cite{Chen1986, Wellstood2008}  and to facilitate analysis of linear and nonlinear laser spectra of molecules in condensed phases.\cite{Mukamel95,TaniIshi09}  Almost independently, theories in the philosophical category have been developed mainly on the basis of the Brownian oscillator model with the aim of understanding how time  irreversibility appears in system dynamics, why macroscopic systems can be treated with classical mechanics instead of quantum mechanics, how wave functions collapse as a result of measurements made with macroscopic instruments, and why and how quantum systems approach a thermal equilibrium state through  interaction with their environments.\cite{Davies76,Spohn80,CaldeiraPRL81,CLAnnlPhys1983,WaxmanLeggett1985,Leggett1987SpinBoson,KuboToda85,GrabertZPhys84,GrabertPR88,Weiss08,Breuer02, Binder19}

While the possibility of treating such problems through analytical approaches is limited in the case of a Brownian-based system, researchers in particular in atomic spectroscopy \cite{Seragent91} and NMR spectroscopy\cite{Guenget2013NMR}  have  used a reduced equation of motion for  density matrix elements, assuming that the  system--bath interaction is weak (perturbative treatment) and that the correlation time of the  bath fluctuations is very short (Markovian assumption). The most commonly used approaches for this kind of problem are the Redfield equation\cite{Redfield65} and the quantum master equation.\cite{Lamb67,Mollow69} It has been shown, however, that these equations do not satisfy the necessary positivity condition without the imposition of a rotating wave approximation (or a secular approximation).\cite{Davies76,Spohn80,Pechukas94,Romero04} Because such approximations modify the form of the system--bath interaction,\cite{Frigerio81,Frigerio85,Pletyukhov2019} the thermal equilibrium state and the dynamics of the original total Hamiltonian are altered.\cite{TanimuraJPSJ06,Tanimura2014,Tanimura2015} In addition to these Markovian equations, phenomenological stochastic approaches have often been employed to account for the non-Markovian dephasing effects of line shape.\cite{Kubo69, Silbey12, Kampen81} These approaches, however, ignore the effects of  dissipation and are formally equivalent to assuming  that the bath temperature is infinite.\cite{TanimuraJPSJ06}

Since the 1980s, when it became possible to investigate  the ultrafast dynamics of molecular motion  using nonlinear laser spectroscopic techniques such as pump--probe and photon echo measurements,  the importance has been realized of non-Markovian effects of the environment, in which the noise correlation time of the environment is comparable to the time scale of the system dynamics. \cite{Mukamel95} Moreover, in many chemical physics and biochemical physics problems, the environment is complex and strongly coupled to the system at finite temperature. Thus, as well as the Markovian approximation, other approximations also become invalid, including the rotating wave approximation, the factorization assumption, and perturbative expansions.\cite{TanimuraJPSJ06} Thus, a great deal of effort has been dedicated to studying the problems of open quantum dynamics with a nonperturbative and non-Markovian system--bath interaction. 

The modified Redfield \cite{MdRedfield98,MdRedfield02} and time-convolution-less (TCL) Redfield equations\cite{TCLshibata77,TCLshibata79,TCLshibata10} are reduced equations of motion that can be derived from the quantum Liouville equation by reducing the heat bath degrees of freedom. Although these approximate approaches are handy for studying problems of open quantum dynamics, their range of validity is limited. 
For example, they are not applicable to a system subject to a time-dependent external force, because the energy eigenstates of the system incorporated in these formalisms are altered in time by the external perturbation. 

The pseudomode approach\cite{Garraway1997,Garraway2009} and the reaction coordinate mapping approach\cite{Nszir2014,Strasberg2018,Strasberg2018B} consider equations of motion that utilize a kind of effective modes whose dynamics are described by the Markovian master equation. While these approaches have wider applicability than conventional reduced equation of motion approaches, the description of long-time behavior that they provide may not be accurate, in particular at very low temperatures, because the Markovian master equation cannot predict the correct thermal equilibrium state. This limitation can be relaxed by introducing a pseudo-Matsubara mode, as in the case of the HEOM formalism.\cite{Nori2019} 

Several variational approaches, such as the multiconfigurational time-dependent Hartree (MCTDH) approach\cite{ML-MCTDH1,ML-MCTDH2, WangTHoss2} and the time-dependent Davydov ansatz (TDDA),\cite{TDDA1,TDDA3,TDDA4} and  asymptotic approaches, such as the effective-mode (EM) approach,\cite{Burghardt1, Burghardt3} the density matrix renormalization group (DMRG),\cite{White1, White2} and the time-evolving density matrix using orthogonal polynomials algorithm (TEDOPA), \cite{AlexPlenio,Plenio2010} have been developed on the basis of a wave function formalism for the total system. The variational approaches can be used to treat nonlinear system--bath coupling, anharmonic bath modes,\cite{WangTHoss2} and a variety of Hamiltonians, for example, the Holstein Hamiltonian;\cite{TDDA1,TDDA3,TDDA4} however, because the bath is described as a finite number of oscillators, the number of bath modes must be increased until convergence is realized to obtain accurate results. This implies that the study of long-time behavior using these approaches requires an intensive computational effort, whereas a reduced equation of motion approach requires a numerical effort that scales only linearly with the simulation time.
Strictly speaking, these wave-function-based approaches can describe only time-reversible processes, and thus, within these approaches, there exists no thermal equilibrium state. Moreover, the inclusion of time-dependent external forces is not as straightforward in these approaches, because the energy of the total system changes owing to the presence of an external force if the perturbation is strong. However, in practice, this kind of approach has wider applicability than any reduced equation of motion approach. 

As  theories of open quantum dynamics,  on the basis of the formally exact path integral or the formally exact reduced equation of motion formalism, numerically  ``exact'' approaches have been developed  that are not subject to the  limitations of the approaches discussed above. Here,  numerically ``exact'' indicates the ability to calculate dynamical and thermal aspects of a reduced system with any desired accuracy that can be clearly verified through non-Markovian tests on the basis of  exact analytical solutions.

The path-integral approaches, most notably  path-integral Monte Carlo, are computationally intensive, because the number of paths to be evaluated grows rapidly with time, while sampling often fails, owing to  phase cancellation of wave functions.\cite{EggerMak94, CaoVoth96} Much effort has been expended in attempting to overcome these problems, for example, by using the quasi-adiabatic path-integral (QUAPI) algorithm,\cite{Makri95, Makri96, Makri96B,Thorwart00,Makri07,JadhaoMakri08,MakriJCP2014,Reichman2010} and to extend the applicability of the path-integral-based methods. Because these approaches can easily incorporate a semiclassical approximation for the bath\cite{Makri2015rev} or introduce a modular scheme to effectively separate the system part and the environmental part,\cite{MakriJCP2018} they have advantages for the study of polyatomic systems treated in multidimensional coordinates. 

Several equation of motion approaches that explicitly handle time-dependent random variables for the heat bath have also been developed.\cite{Strunz97,Stockburger99,Stockburger02,Shao04} These approaches are formally exact, but  realization of the random variables in the low-temperature regime is difficult, and the applicability of these approaches is still limited to simple systems.

The reduced hierarchical equations of motion (HEOM) theory is a method that can describe the dynamics of a system with a nonperturbative and non-Markovian system--bath interaction at finite temperature, even under strong time-dependent perturbations.\cite{TanimuraJPSJ06,Tanimura89A,TanimuraPRA90,TanimuraPRA91,TanimuraJCP92,IshizakiJPSJ05,Tanimura2014,Tanimura2015,Xu05,Yan06} In this formalism, the effects of higher-order non-Markovian system--bath interactions are mapped into the hierarchical elements of the reduced density matrix. This formalism is valuable because it can be used to treat not only strong system--bath coupling, but also quantum coherence or quantum entanglement between the system and bath (``bathentanglement''), in particular for a system subject to a time-dependent external force and nonlinear response functions.\cite{DijkstraPRL10,DijkstraJPSJ12entangle, DijkstraPTRS2012nonMarkov} Although the HEOM approach is numerically very expensive in comparison with the other reduced equation of motion approaches, a variety of analytical and numerical techniques can be employed to integrate the HEOM. With these features, the HEOM approach offers wide applicability. 
For example, it is possible to study quantum heat-engine and quantum ratchet problems, in which the nonequilibrium steady state is described respectively by the long-time behavior of a system strongly coupled with two heat baths at different temperatures and by the long-time behavior of a system under periodic time-dependent external forces, as illustrated in the following sections.  
In this article, we present an overview of the HEOM theory, focusing on its theoretical background and applications to help further the development of  investigations in this field. 

The organization of the remainder of the paper is as follows. In Sec.~\ref{sec:System}, we present typical model systems for open quantum dynamics. In Sec.~\ref{sec:HEOM}, we explain the standard HEOM and their characteristic features. In Sec.~\ref{sec:nonMarkovinTests}, we demonstrate the accuracy of the HEOM by numerically  ``exact'' tests. In Sec.~\ref{sec:Evolution}, we illustrate the variety of HEOM for various systems. In Sec.~\ref{sec:Applications}, we review  various applications of the HEOM theory. We present future perspectives for HEOM theory  in Sec.~\ref{sec:Future}.

\section{SYSTEM}
\label{sec:System}
We consider a situation in which a primary system interacts with heat baths that give rise to dissipation and fluctuation in the system. \cite{CaldeiraPRL81,CLAnnlPhys1983,WaxmanLeggett1985,GrabertPR88,GrabertZPhys84,Leggett1987SpinBoson,KuboToda85,Weiss08,Breuer02} The total Hamiltonian is expressed as
\begin{align}
\hat H_\mathrm{tot} = \hat H_A + \hat H_{I + B},
\label{eq:Spinboson}
\end{align}
where $\hat H_A$ is the Hamiltonian of the system and $\hat H_{I + B}$ is the bath Hamiltonian, which includes the system--bath interaction.
The bath degrees of freedom (labeled by $a$) are treated as $N_a$  harmonic oscillators: 
\begin{align}
\hat H_{I + B} =  \sum_{a} \sum_{j=1}^{N_{a}} \left[ \frac{(\hat p_j^{a})^2 }{2m_j^{a} } + \frac{1}{2}m_j^{a} (\omega _j^{a})^2 (\hat x_j^{a})^2  - \alpha_j^{a} \hat V^{a} \hat x_j^{a} \right] ,
\label{eq:bath}
\end{align}
where the momentum, position, mass, and frequency of the $j$th oscillator in the $a$th bath are given by $\hat{p}_{j}^{a}$, $\hat{x}_{j}^{a}$, $m_{j}^{a}$, and $\omega_{j}^{a}$, respectively,  $\hat V^{a}$ is the system part of the interaction, and $\alpha_j^{a}$ is the coupling constant between the system and the $j$th oscillator in the $a$th bath.
The above bath model is commonly applied to systems possessing discretized energy spaces, for which it is expressed as 
$\hat H_{A} = \sum _{j} \hbar \omega_j | j \rangle\langle j | + \sum _{j\ne k} \hbar \Delta_{jk}(t)| j \rangle \langle k|$ and $\hat V^{a}= \sum _{j, k} V_{jk}^{a} | j \rangle\langle k |$, which are defined using the bra and ket vectors of the $j$th eigenenergy states of the system  $| j \rangle$ and $\langle j |$. A notable application of this kind is to the spin-boson Hamiltonian for $j= 0$ or $1$.\cite{Leggett1987SpinBoson,KuboToda85,Weiss08,Breuer02} The model is also commonly applied to systems possessing continuous configuration spaces,  for which it is expressed in general form as
$\hat H_{A} = {\hat {\bf p}^2}/{2m} + U( \hat {\bf q}; t)$, for a system with  mass $m$ and  potentials $U(\hat {\bf q})$ and $\hat V^{a}=\hat{V}^{a}(\hat {\bf p}, \hat {\bf q})$ described in terms of the multidimensional momentum $\hat {\bf p}$ and coordinate $\hat {\bf q}$.\cite{CaldeiraPRL81,CLAnnlPhys1983,WaxmanLeggett1985,GrabertZPhys84,GrabertPR88} A notable example of such an application is to the Brownian Hamiltonian.  
 In the Brownian case,  to avoid an unphysical energy divergence, a counterterm $\sum_{a}\sum_{j} {(\alpha_{j}^{a}V^{a})^2}/{2m_{j}^a(\omega _{j}^{a})^{2}}$ is introduced in the bath Hamiltonian such that the replacement $\hat{x}_{j}^{a} \rightarrow \hat{\bar {x}}_{j}^{a} \equiv  \hat{x}_{j}^{a} - {\alpha_{j}^{a} \hat{V}^{a}}/{m_{j}^{a}(\omega_{j}^{a})^2} $ is made \cite{CaldeiraPRL81,CLAnnlPhys1983} to maintain  translational symmetry in the case $U(\hat {\bf q})=0$.\cite{TanimuraPRA91,TanimuraJPSJ06} In  atomic spectroscopy, this divergence phenomenon is known as the Lamb shift.\cite{Lamb67,Mollow69}
 
 In many physical processes, molecular motion and electronic excitation states are coupled and play important roles simultaneously.
Using this bath model, we can further include the effects of the electronically excited states by extending the space of the system Hamiltonian as
$\hat H_{A} = \sum _{j}  | j \rangle ({\hat {\bf p}^2}/{2m}) \langle j | + \sum_{j, k} | j \rangle U_{jk}( \hat {\bf q};t)\langle k|$.\cite{Mukamel95,YangTonuOliverRev2015}

With the features described above, the Brownian heat bath possesses wide applicability, despite its simplicity. This is because the influence of the environment can in many cases be approximated by a Gaussian process, owing to the cumulative effect of a large number of weak environmental interactions, in which case the ordinary central limit theorem is applicable,\cite{Kampen81} while the distribution function of the harmonic oscillator bath itself also exhibits a Gaussian distribution.\cite{TanimuraJPSJ06, Kampen81}

Hereinafter, we assume that each bath is independent of the others.
The $a$th heat bath can be characterized by the spectral distribution function (SDF), defined by
\begin{align}
  J_a (\omega) \equiv \sum_{j=1}^{N_{a}}\frac{\hbar (\alpha_{j}^a)^2}{2m_{j}^a \omega_{j}^a } \delta(\omega-\omega_{j}^a),
  \label{eq:J_wgeneral}
\end{align}
and the inverse temperature $\beta \equiv 1/k_{\mathrm{B}}T$, where $k_\mathrm{B}$ is the Boltzmann constant. Various   environments, for example, those consisting of nanostructured materials, solvents, and/or protein molecules, can be modeled by adjusting the form of the SDF. 
For the heat bath to be an unlimited heat source  possessing an infinite heat capacity, the number of heat bath oscillators $N_a$ is effectively made infinitely large by replacing $J_a (\omega)$ with a continuous distribution. When we reduce the bath degrees of freedom to have the reduced density matrix ${\hat \rho}_A (t)=\mathrm{tr}_B\{ {\hat \rho}_{A+B} (t)\}$, the bath acts as a noise source for the system via the system--bath interaction.\cite{Tanimura89A,TanimuraJPSJ06} The random variable defined as $\hat \Omega(t)\equiv \sum_j \alpha_j  \hat x_j$ describes the thermalization of the system through  fluctuation and dissipation. In the case of a harmonic bath, the contributions of higher-order cumulants vanish, and the effects of dissipation are expressed in terms of the response function defined as $L_1 (t) = i\langle \hat \Omega (t) \hat \Omega -\hat \Omega  \hat \Omega (t) \rangle_B/\hbar$, while those of thermal fluctuations are  expressed as $L_2(t) =  \langle \hat \Omega(t)\hat \Omega + \hat \Omega \hat \Omega (t) \rangle_B/2$, where $\langle \cdots \rangle_{B}$ represents the thermal average of the bath degrees of freedom.\cite{Leggett1987SpinBoson,TanimuraJPSJ06,Tanimura2014,Tanimura2015} The thermal equilibrium state of the system is realized through the energy exchange arising from fluctuation and dissipation: the condition in which there is a thermal equilibrium state is completely described by the quantum version of the fluctuation--dissipation theorem, because of the Gaussian nature of the noise, which is equivalent to assuming a harmonic heat bath.\cite{Tanimura89A,KuboToda85}

\section{REDUCED HIERARCHICAL EQUATIONS OF MOTION}
\label{sec:HEOM}
To illustrate the characteristic features of the HEOM formalism in a simple manner, in this section, we consider a Drude SDF (an Ohmic distribution with a Lorentzian cutoff) for a single bath model, expressed  as\cite{TanimuraJPSJ06,Tanimura89A,TanimuraPRA90,TanimuraPRA91,TanimuraJCP92,IshizakiJPSJ05,Tanimura2014,Tanimura2015,Xu05,Yan06}
\begin{equation}
 J(\omega) = \frac{\hbar \eta}{\pi}\frac{\gamma^2\omega}{\gamma^2+\omega^2},
\label{JDrude}
\end{equation}
where the constant $\gamma$ represents the width of the spectral distribution of the collective bath modes and is the reciprocal of the relaxation time of the noise induced by the bath. The parameter $\eta$ is the system--bath coupling strength, which represents the magnitude of fluctuation and dissipation. By taking $\gamma \to \infty$, the Drude form reduces to the Ohmic form $J(\omega)= \hbar \eta \omega/\pi$, which is often employed to obtain Markovian noise in the high-temperature case. 

The relaxation function is temperature-independent and is expressed as $L_1(t) = {\bar c}_0
e^{ - \gamma | t |}$, with ${\bar c}_0=\hbar \eta \gamma^2$, while the noise correlation is given by
$L_2(t) = c_0 e ^{ - \gamma |t|}  + \sum_{k}  {c_k} e ^{ - \nu _k |t|}$, with 
$c_0 = \hbar \eta\gamma^2 \cot (\beta\hbar\gamma/2)/2$ and $c_k  = - 2\eta \gamma ^2{\nu_k}/\beta(\gamma^2  - \nu_k^2)$ for the $k$th Matsubara frequency $\nu_k \equiv 2 \pi k /\beta \hbar$.\cite{TanimuraJPSJ06,Tanimura89A,TanimuraPRA90,TanimuraPRA91,TanimuraJCP92,IshizakiJPSJ05}

\subsection{The positivity condition}
\begin{figure}
\includegraphics[width=0.5\columnwidth]{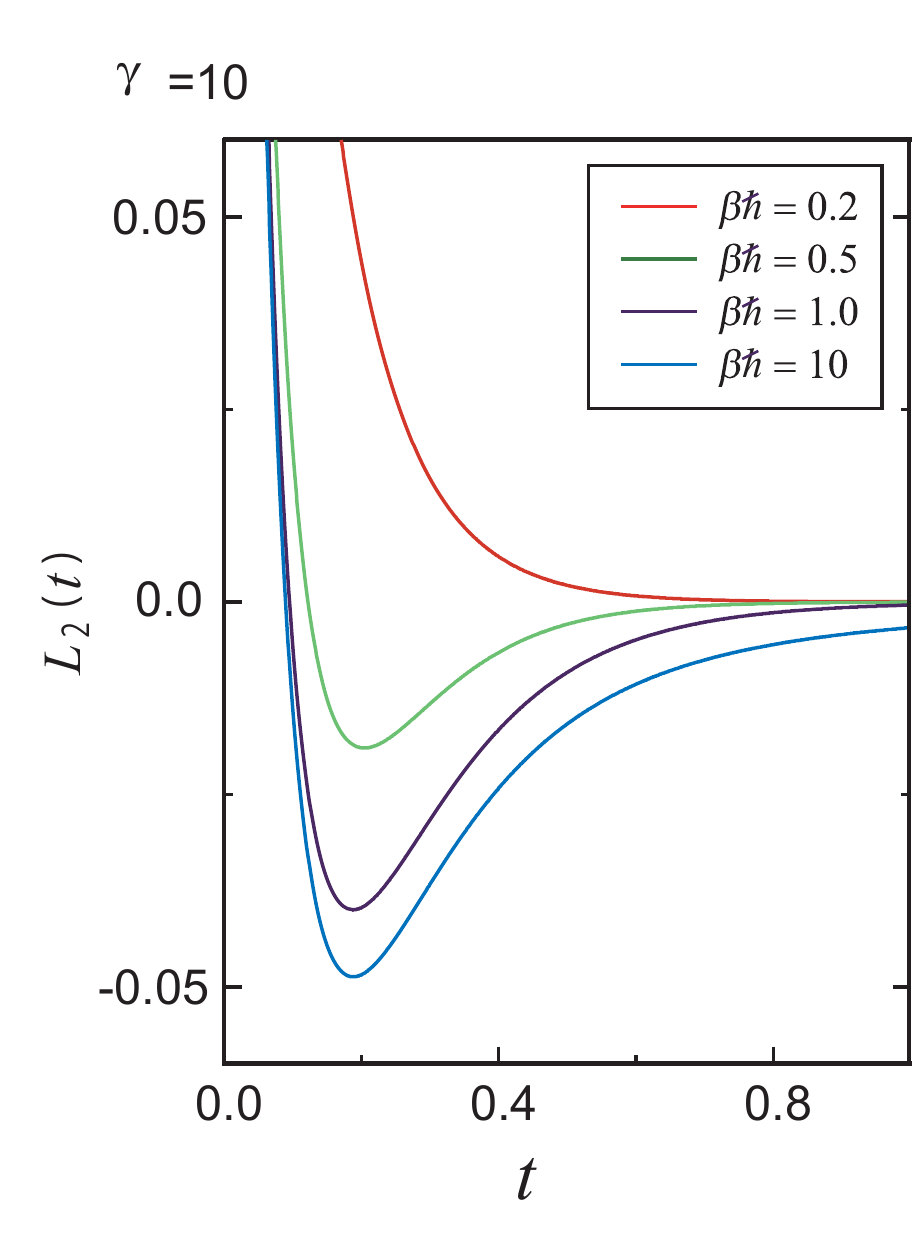}
\caption{\label{L_2Fig}Noise correlation (fluctuation) function $L_2(t)$ for the Drude spectral distribution, Eq.~\eqref{JDrude}, depicted as a function of the dimensionless time $t$ in the fast-modulation case ($\gamma=10$) for the different inverse temperatures $\beta \hbar$. The profile of $L_2(t)$ in the high-temperature case becomes similar to that of the response function $L_1(t)$. While $L_2(t)$ exhibits a Markovian nature in the Ohmic case ($\gamma \to \infty$) at high temperature, it becomes negative at low temperature (large $\beta\hbar$ ) owing to the contribution of the Matsubara frequency terms. Because of this feature, the quantum noise cannot be Markovian. Y. Tanimura, \href{https://doi.org/10.1143/JPSJ.75.082001} {J. Phys. Soc. Jpn.} {\bf 75}, 082001 (2006); licensed under a Creative Commons Attribution (CC BY) license.}
\label{fig:L2}
\end{figure}

The Markovian assumption is commonly employed in quantum open dynamics. Because the Markovian assumption is unrealistic in the physical problems, owing to the ignorance of the non-Markovian features of quantum fluctuations, we encounter a breakdown of the positivity condition of the population states if we do not employ further approximations, most notably the rotating wave approximation (RWA) (or a secular or resonant approximation).\cite{Davies76,Spohn80,Pechukas94,Romero04,Frigerio81,Frigerio85,Pletyukhov2019}  To illustrate this point, we plot $L_2(t)$ in Drude cases for large $\gamma$ in Fig.~\ref{L_2Fig}.\cite{TanimuraJPSJ06,Tanimura2015} The profile of the response function $L_1 (t)$ is similar to that of $L_2 (t)$ in the high-temperature case and becomes a $\delta$ function [$\propto \delta(t)$] as $\gamma \to \infty$. At low temperatures, however, the noise correlation $L_2(t)$ becomes negative owing to the contribution from terms with  $c_k e^{- \nu_k t}$ for small $k$ in the region of small $t$. This behavior, which we call  low-temperature-induced non-Markovianity, is a characteristic feature of  quantum thermal noise.\cite{TanimuraJPSJ06,Tanimura2015}
This indicates that the validity of the Markovian assumption in the quantum case is limited only in the high-temperature regime. Approaches that employ the Markovian master equation and the Redfield equation, which are usually applied to systems possessing discretized energy states, ignore or simplify such non-Markovian features of  fluctuations, and this is the reason why the positivity condition is broken.\cite{TanimuraJPSJ06,Tanimura2014,Tanimura2015} As a way to resolve this problem, the RWA is often employed,\cite{Frigerio81,Frigerio85,Pletyukhov2019} but a system treated under this approximation will not satisfy the fluctuation--dissipation theorem. Thus the use of such an approximation may introduce significant errors in the thermal equilibrium state and in the time evolution of the system toward equilibrium, as will be illustrated in Sec.~\ref{sec:nonMarkovinTests}. In the classical limit, with $\hbar$ tending to zero, $L_2(t)$ is always positive.

As can be seen from the form of $L_1(t)$ and $L_2(t)$, there are two types of non-Markovian components, as illustrated in the Drude case: one is of  mechanical origin and is characterized by  dissipation [i.e., $L_{1}(t)$] and fluctuations [i.e., $L_{2}(t)$] expressed in terms of $e^{-\gamma t}$, and the other is of  quantum thermal origin and is characterized by  fluctuations only [i.e., $L_{2}(t)$], with the functions expressed in terms of $c_ke^{-\nu_k t}$ for $k\ge 1$, as the definition of $\nu_k=2\pi k/\beta\hbar$ indicates. When $\gamma$ is much larger than $\nu_1$, the mechanical contribution with $e^{-\gamma t}$ vanishes for $t>1/\nu_1$, and then the effects from the quantum thermal noise arise. The quantum thermal fluctuations exhibit peculiar behavior  compared with the mechanical fluctuations, because the amplitude of the noise becomes negative for  large $\gamma$, as illustrated in Fig.~\ref{fig:L2}. While the mechanical noise term contributes to the relaxation and dephasing processes through  fluctuation and dissipation, the quantum thermal noise term contributes to the dephasing process. This difference can be observed even in the free induction decay  of a spin-boson system, if $\gamma$ is large and the temperature is very low.\cite{TakahashiJPSJ2020}

\subsection{HEOM for density operators}
\begin{figure}
\includegraphics[width=\columnwidth]{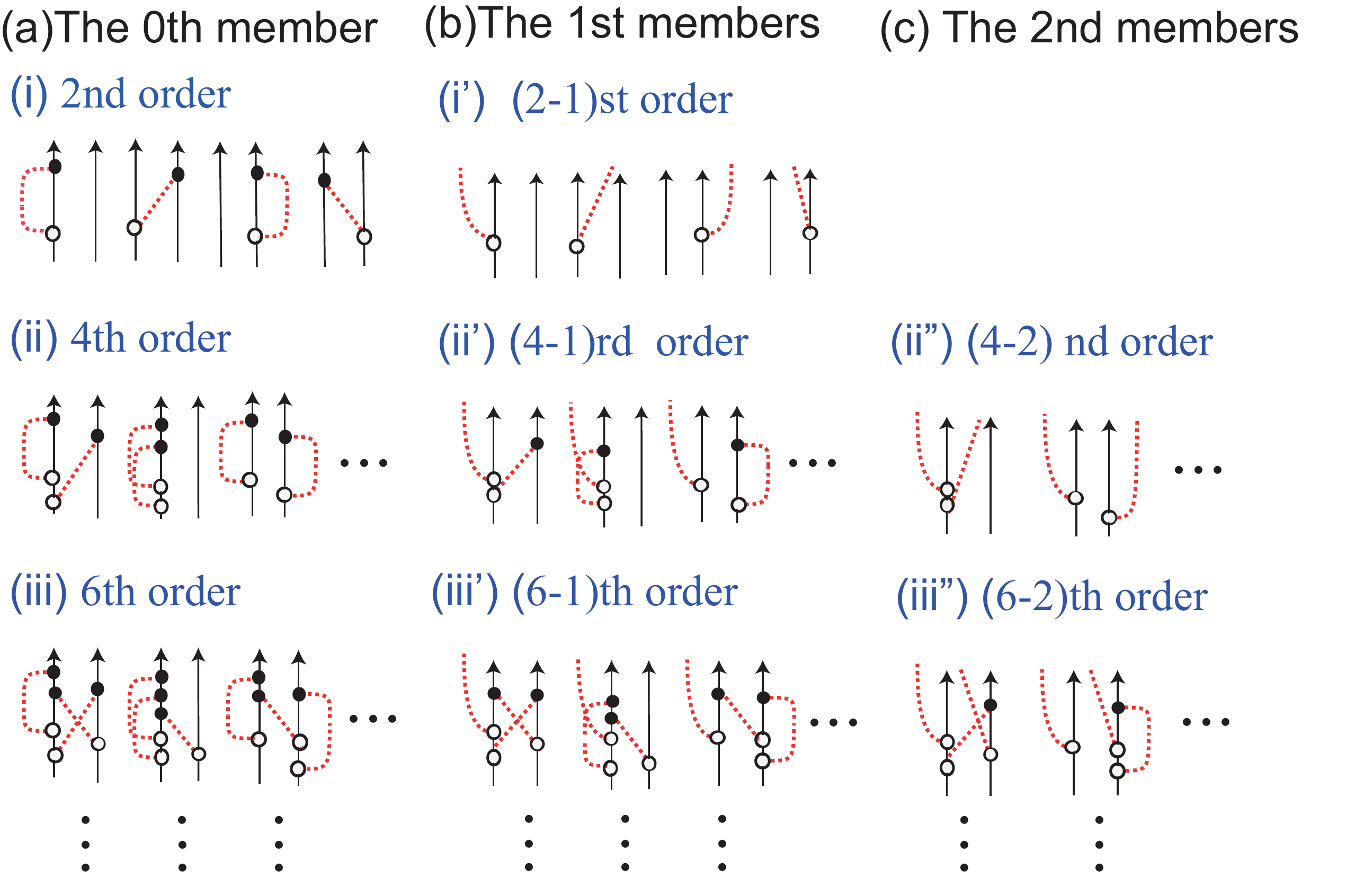}
\caption{\label{dsidedFyenman}Double-sided Feynman diagrams of density matrices 
with the three lowest system--bath interactions.
(a-i)--(a-iii) depict the elements of  $\hat \rho _{0,0, \dots ,0}^{(0)}(t)$ in terms of the second-,  fourth-, and sixth-order system--bath interactions, whereas (b-i$'$)--(b-iii$'$) and (c-ii$''$)--(c-iii$''$) depict the elements of the first and second members of the hierarchy, which involve $\hat \rho _{0,0, \dots ,0}^{(1)}(t)$ and $\hat \rho _{0,0, \dots ,0}^{(2)}(t)$, etc.
In each diagram, the left and right solid lines represent the time evolution of the system, and the red dashed lines represent the bath excitations.
The white and black circles correspond to the system--bath interactions involved in $\hat \Phi$ and $\hat \Theta _j$ ($j=0, 1, \dots $), respectively. Here, the quantum entanglement between the system and the bath (bathentanglement) is illustrated by the red dashed curves. A detailed explanation can be found in Refs.~\onlinecite{TanimuraJPSJ06,Tanimura2014,Tanimura2015}. 
}
\end{figure}

The HEOM are derived by differentiating the reduced density matrix ${\hat \rho}_A (t)$, defined by a path integral\cite{TanimuraJPSJ06, Tanimura2014,Tanimura2015,Tanimura89A,TanimuraPRA90, IshizakiJPSJ05, TanimuraPRA91} or by time-ordered operators with the application of Wick's theorem.\cite{Ishizaki2009B} The HEOM can describe the dynamics of the system under nonperturbative and non-Markovian system--bath interactions with any desired accuracy at finite temperature.\cite{TanimuraJPSJ06, Tanimura2014,Tanimura2015} In their original formulation, these equations of motion were limited to the case in which the SDF takes the Drude form and the bath temperature is high.\cite{Tanimura89A} Subsequently, these  limitations have been removed, and the equations are applicable to  arbitrary SDFs and temperatures.\cite{TanimuraPRA90,TanimuraPRA91} 

While conventional approaches employing reduced equations of motion eliminate the bath degrees of freedom completely, the HEOM approach retains information with regard to bath dynamics, including  system--bath coherence or system--bath entanglement (bathentanglement) in the hierarchy elements $\hat \rho _{{j_1}, \dots ,{j_K}}^{(n)}$, where the set of indices $n$ and $\{j_k\}$ arise from the hierarchical expansion of the noise correlation functions in terms of $e^{- \gamma  t}$ and $e^{- \nu_k t}$, respectively. \cite{TanimuraJPSJ06, Tanimura2014,Tanimura2015}
The 0th element is identical to the reduced density matrix, ${\hat \rho}_A (t)= \hat \rho _{0,0, \dots ,0}^{(0)}(t)$, which includes all orders of the system--bath interactions, and it is the exact solution of the Hamiltonian, Eq.~(\ref{eq:Spinboson}).
For the set of $n$ and $\{ j_k\}$ with 
$N=\omega_0/\min(\gamma, {\nu_1}) > (n+\Sigma_{k=1}^K j_k)$, where $\omega_0$ is the characteristic frequency of the system and $K$ is a cutoff  satisfying $K \gg \omega_0 /\nu_1$, the HEOM are expressed as\cite{TanimuraJPSJ06,Tanimura89A,TanimuraPRA90,TanimuraPRA91,Tanimura2014,IshizakiJPSJ05}
\begin{widetext}
\begin{align}
	\frac{\partial}{\partial t}
	\hat{\rho}_{j_1,\dots,j_K}^{(n)}(t)
	={}&-\left[
	  i\hat{ L}_A+ n \gamma+
		\sum_{k=1}^K j_k\nu_k
		+\hat{\Xi}
	\right]\!
	\hat{\rho}_{j_1, \dots,j_K}^{(n)}(t)
      +\hat{\Phi}\!\!	\left[ \hat{\rho}_{j_1, \dots,j_K}^{(n+1)}(t) +\sum_{k=1}^K 
	\hat{\rho}_{j_1, \dots,j_k+1,\dots,j_K}^{(n)}(t) \right] \notag\\[6pt]
	& +  n {\hat \Theta_0}
	\hat{\rho}_{j_1,\dots,j_K}^{(n-1)} (t)	
	+ \sum_{k=1}^K 
j_k	 {\hat \Theta_k}	\hat{\rho}_{j_1,\dots,j_k-1,\dots,j_K}^{(n)}(t),
\label{eq:HEOM}
\end{align}
\end{widetext}
where $\hat{L}_A$ is the Liouvillian of $\hat H_A$, defined as $\hat{L}_A\equiv \hat H_A ^{\times}/\hbar$,
\[
\hat{\Xi}=-\left( \sum_{k=K+1}^{\infty} \frac{c_k}{\nu_k}\right) \frac{\hat{V}^\times\hat{V}^\times }{\hbar^2} 
\]
 is the renormalization operator to compensate for the effects of the cutoff $K$, and 
 \[
 \hat{\Phi}=\frac{-i {\hat V}^\times }{\hbar}, \ \ {\hat \Theta}_0=\frac{-i(c_0 {\hat V}^\times - i{\bar c}_0 {\hat V}^\circ)}{\hbar}, \ \  
 {\hat \Theta}_k=\frac{-ic_k {\hat V}^\times }{\hbar}
 \]
for an integer $k \ge 1$.
Here, the hyper-operators $\hat{\mathcal{O}}^\times\hat{f}	\equiv	\hat{\mathcal{O}} \hat{f}-\hat{f}\hat{\mathcal{O}}$
and 
$\hat{\mathcal{O}}^\circ\hat{f}	\equiv \hat{\mathcal{O}}\hat{f}+\hat{f}\hat{\mathcal{O}}$
for any operator $\mathcal{\hat{O}}$ and operand $\hat{f}$ are those introduced by Kubo{\cite{Takagahara}} and Tanimura, \cite{Tanimura89A} respectively.

\begin{figure}[t]
\includegraphics[width=0.65\columnwidth]{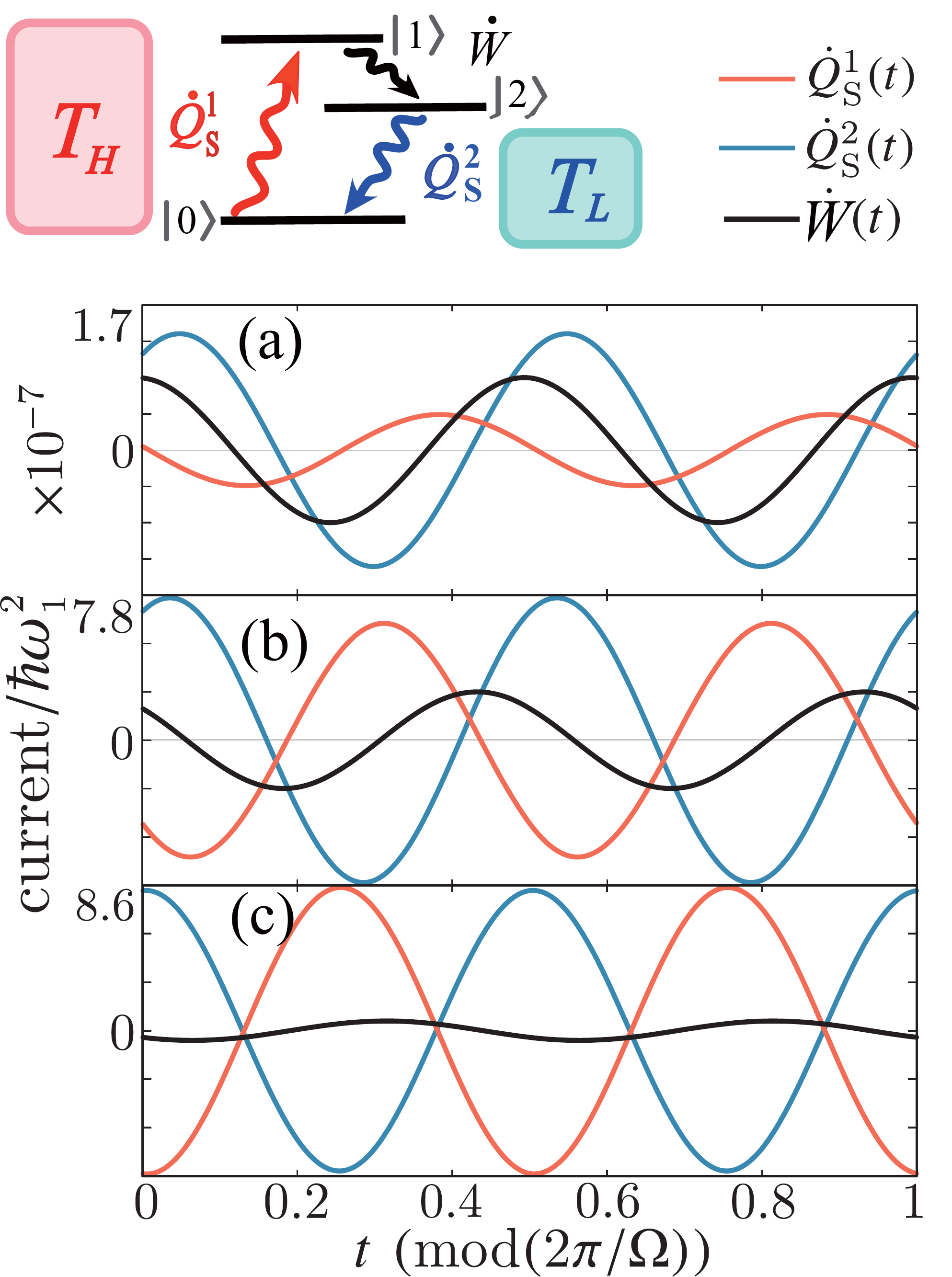}
\caption{\label{heatengine:g}The three-level heat engine model considered here consists of three states, denoted by $| 0 \rangle$, $| 1 \rangle$, and $| 2 \rangle$, coupled to two bosonic baths. The system is driven by a periodic external field with frequency $\Omega$. This model acts as a quantum heat engine when population inversion between the two excited states $|1\rangle$ and $|2\rangle$ occurs.
We illustrate the time dependences of the heat current 
for $\eta_1$ = (a) 0.01, (b) 0.1, and (c) 1 with $\eta_2 = 0.001$ defined from the system energy, $\dot{Q}_\mathrm{S}^1$ and $\dot{Q}_\mathrm{S}^2$, and the power, $\dot W$,that arise from transitions under one cycle of the external force. The time delays observed for the heat currents imply that the transition $|0\rangle \to |1\rangle \to |2\rangle$ is cyclic. This behavior can be regarded as a microscopic manifestation of a quantum heat engine. Details are presented in Refs.~\onlinecite{Katoengine2016,Katobook2019}.
 Reproduced from A. Kato and Y. Tanimura, \href{https://doi.org/10.1063/1.4971370} {J. Chem. Phys.}{\bf 145}, 224105 (2016), with the permission of AIP Publishing.
}
\end{figure}

In this formalism, the effects of the higher-order non-Markovian system--bath interactions are mapped into the hierarchical elements of the reduced density matrix (see Fig.~\ref{dsidedFyenman}).
The operator $\hat{\Phi}$ demolishes the bath states from $\hat{\rho}_{j_1, \dots,j_K}^{(n+1)}(t)$ and $\hat{\rho}_{j_1, \dots,j_k+1,\dots,j_K}^{(n)}(t)$ to $\hat{\rho}_{j_1,\dots,j_K}^{(n)}(t)$, whereas $\hat{\Theta}_0$ and $\hat{\Theta}_k$ create the bath states from $\hat{\rho}_{j_1,\dots,j_K}^{(n-1)} (t)$ and $\hat{\rho}_{j_1,\dots,j_k-1,\dots,j_K}^{(n)}(t)$ to $\hat{\rho}_{j_1,\dots,j_K}^{(n)}(t)$, respectively;  $\hat{\rho}_{j_1, \dots,j_K}^{(n\mp 1)}(t)$ and $\hat{\rho}_{j_1, \dots,j_k\mp 1,\dots,j_K}^{(n)}(t)$ involve the first-order higher ($-$) and lower ($+$) interactions   
than  $\hat{\rho}_{j_1,\dots,j_K}^{(n)}(t)$. 
Thus, one can regard Eq.~\eqref{eq:HEOM} as a type of rate law among the density operators with three different bath-excited states, $\hat{\rho}_{j_1, \dots,j_K}^{(n\pm 1)}(t)$ and $\hat{\rho}_{j_1, \dots,j_k \pm,\dots,j_K}^{(n)}(t)$; the time evolution of $\hat{\rho}_{j_1,\dots,j_K}^{(n)}(t)$ is determined by its time evolution operator $(\hat{ L}_A+ n \gamma+\sum_{k=1}^K j_k\nu_k+\hat{\Xi})$ and the incoming and outgoing contributions of the hierarchical elements connected through  $\hat{\Phi}$, $\hat{\Theta}_0$, and $\hat{\Theta}_k$.\cite{TanimuraJPSJ06} The HEOM approach can treat the noise correlation even if it becomes negative (as shown in Fig.~\ref{L_2Fig}) in a nonperturbative and non-Markovian manner, because this approach retains the information of the system--bath correlation in the hierarchy elements with the set of indices $n$ and $\{j_k\}$. Moreover, these hierarchical elements allow us to evaluate the bathentanglement, and were utilized to calculate the heat flow between the system and the bath.\cite{Katoflow2015,Katoengine2016,Katobook2019, Shi2012SBcorre}

The HEOM approach differs conceptually  from conventional perturbative expansion approaches, where the $0$th member does not include any system--bath interactions and where higher members take into account higher-order system--bath interactions: in the HEOM approach, $\hat \rho _{0,0, \dots ,0}^{(0)}(t)$ includes all orders of the system--bath interactions, and it is the exact solution of the Hamiltonian, Eq.~(\ref{eq:Spinboson}).\cite{TanimuraJPSJ06,Tanimura2014,Tanimura2015, Tanimura89A,TanimuraPRA90,TanimuraPRA91,TanimuraJCP92,IshizakiJPSJ05}
As an example, a numerical solution of the HEOM for a three-level heat engine model is presented in Fig.~\ref{heatengine:g}.\cite{Katoengine2016,Katobook2019}

\subsection{Quantum hierarchical Fokker--Planck equations}
\begin{figure}[t]
\includegraphics[width=0.91\columnwidth]{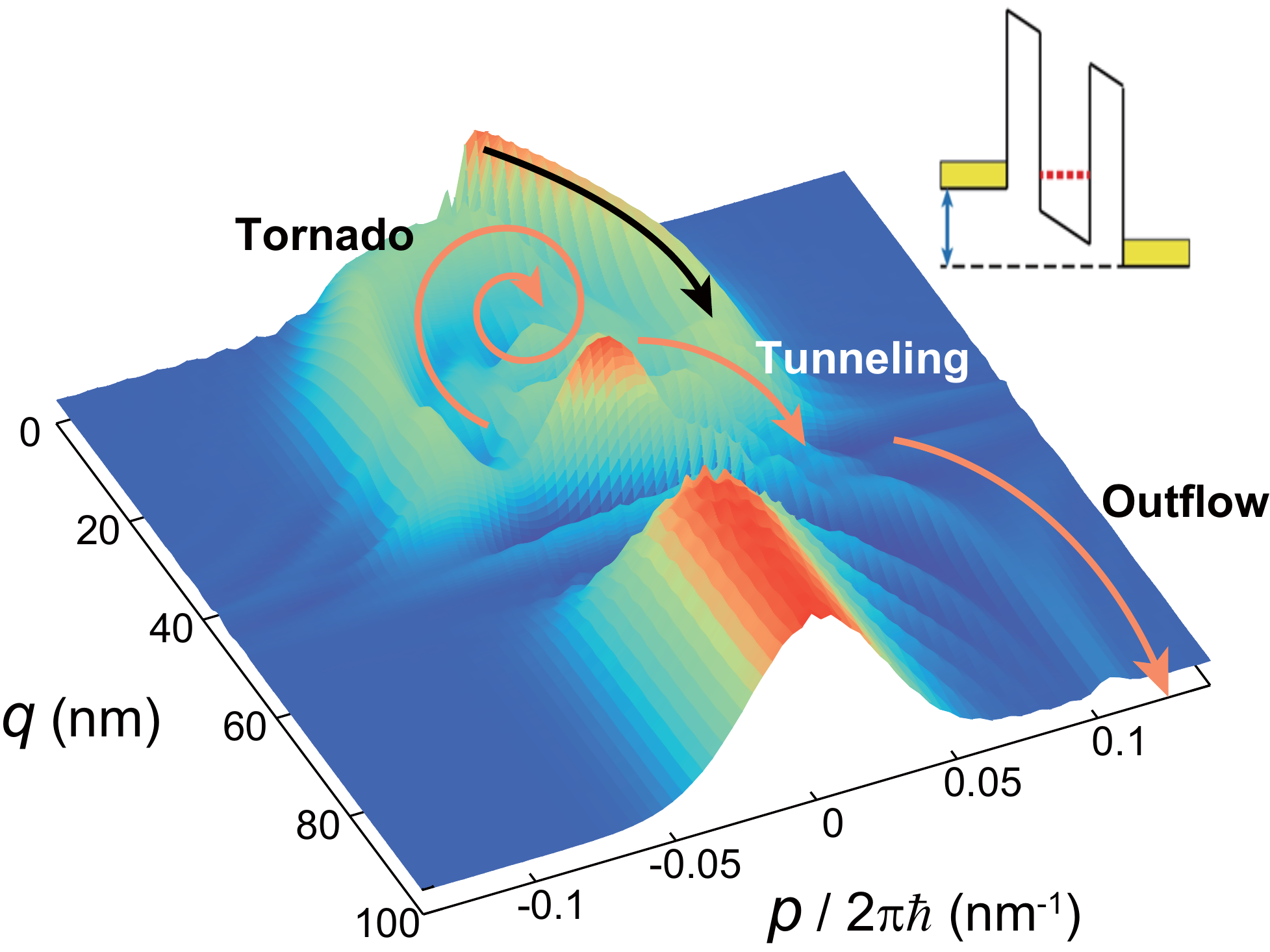}
\caption{\label{RTD:g} A snapshot of the Wigner distribution function for a self-current oscillation of a resonant tunneling diode. Current flows into the system from the emitter side of the boundary ($q=0$~nm).
Then, part of the current is scattered by the emitter side of the barrier.
The scattered electrons flow in a tornado-like manner to the central peak in the emitter basin owing to dissipation. The shaking motion of the effective potential periodically accelerates the component at the central peak to the tunneling state, and the current thus exhibits steady oscillations, although the Hamiltonian is time-independent. Details are presented in Refs.~\onlinecite{SakuraiJPSJ13, SakuraiNJP14,GrossmannJPC14}. A. Sakurai and Y. Tanimura, \href{https://doi.org/10.7566/JPSJ.82.033707} {J. Phys. Soc. Jpn.} {\bf 82}, 033707 (2013); licensed under a Creative Commons Attribution (CC BY) license.}
\end{figure}

For a system described in  coordinate and momentum space,  it is often more convenient to use the phase space representation than the energy eigenstate representation, in particular to implement various boundary conditions, most notably periodic, open, and inflow--outflow boundary conditions, and to take the classical limit to identify pure quantum effects. For the density matrix element $\rho_{j_1,\dots,j_K}^{(n)} (q, q';t)$, we introduce the Wigner distribution function, which is the quantum analog of the classical distribution function in phase space and is defined as \cite{Wigner32, Frensley90}
\begin{multline}
W_{j_1,\dots,j_K}^{(n)}(p,q;t)  \equiv\\  \frac{1}
{{2\pi \hbar }}\int_{ - \infty }^\infty  {dx} \, e^{ipx/\hbar }  \rho_{j_1,\dots,j_K}^{(n)} \!\left( {q - \frac{x}
{2}, q + \frac{x}
{2}};t \right).
\end{multline}
The Wigner distribution function is a real function, in contrast to the complex density matrix. In terms of the Wigner distribution, the quantum Liouvillian for the system Hamiltonian takes the form \cite{Frensley90}
\begin{align}
& - \mathcal{\hat L}_{QM}W_{j_1,\dots,j_K}^{(n)} (p, q) \equiv \notag \\ 
 & \qquad - \frac{p}{m}\frac{\partial }{{\partial q}}W_{j_1,\dots,j_K}^{(n)} (p, q)  \notag \\
&\qquad - \frac{1}{\hbar }\int_{ - \infty }^\infty {\frac{{dp'}}
{{2\pi \hbar }} U_W (p - p', q)} W_{j_1,\dots,j_K}^{(n)} (p', q), 
\label{eq:quantumLioiv}
\end{align}
where 
\begin{multline*}
 U_W (p, q) = \\
 2 \int_{0}^\infty dx \sin 
\!\left( \frac{px}{\hbar}\right ) 
\left[ U \!\left( q + \frac{x}{2}\right) - U\!\left(  q - \frac{x}{2}\right) \right].
\end{multline*}
The quantum hierarchical Fokker--Planck  equations (QHFPE) with  counterterm can then be expressed as\cite{Tanimura2015, SakuraiJPSJ13,SakuraiNJP14,GrossmannJPC14,KatoJPCB13}
\begin{widetext} 
\begin{align}
 \frac{\partial}{\partial t}{W}_\jk^{(n)}(p,q;t) &= -\left( \mathcal{\hat L}_{QM}  + n\gamma + \sum_{k=1}^K j_k\nu_k + \hat{\Xi'} \right) W_\jk^{(n)}(p,q;t) \nonumber \\
	& + \hat{\Phi}' \!\left[ W_\jk^{(n+1)}(p,q;t) + \sum_{k=1}^K W_\jkp^{(n)}(p,q;t) \right] \nonumber \\
	& + n{{\hat {\bar \Theta} }_0} W_\jk^{(n-1)}(p,q;t) 
 + \sum_{k=1}^K j_k \hat{\Theta}_k' W_\jkm^{(n)}(p,q;t),
\label{heom_wig}
\end{align}
\end{widetext}
where the dashed operators are the Wigner representations of the operators in Eq.~\eqref{eq:HEOM}. For linear--linear system--bath coupling, $V(q)=q$, they are expressed as 
\begin{align*}
\hat \Phi' &=  \frac{\partial }{\partial p}, & {\hat {\bar \Theta} }_0 &= { \eta }\left[\frac{p}{m} + c_0 \cot\!\left( \frac{\beta \hbar \gamma}{2}\right) \frac{\partial}{\partial p} \right],\\
 \hat{\Theta}_k'&=c_k \frac {\partial }{\partial p}, & 
\hat{\Xi'} &\equiv -\frac{\eta}{\beta}\left(\sum_{k=K+1}^{\infty} c_k  \right) \frac{\partial^2 }{\partial p^2}. 
\end{align*}

As an example, a numerical solution of the QHFPE for a self-current oscillation of a resonant tunneling diode is presented in Fig.~\ref{RTD:g}.\cite{SakuraiJPSJ13, SakuraiNJP14,GrossmannJPC14} The QHFPE for linear--linear (LL) plus square--linear (SL), $V(q)= q + c q^2$, and  exponential--linear (EL), $V(q)=e^{-cq}$, system--bath coupling cases for any constant $c$ are given in Refs.~\onlinecite{TaniIshiACR09,SteffenTanimura00,TanimuraSteffen00,KatoTanimura02,KatoTanimura04,SakuraiJPC11} and Ref.~\onlinecite{Jianji2020}, respectively. It should be noted that because the counterterm is incorporated into the equations of motion to maintain the translational symmetry of the reduced equations of motion, the QHFPE are not merely a Wigner transformation of the regular HEOM, Eq.~\eqref{eq:HEOM}.

In the LL high-temperature case, we can ignore the low-temperature correction terms, and the above equations for $W^{(n)} (p, q)=W_{0,\dots,0}^{(n)} (p, q)$ further reduce to \cite{TanimuraPRA91,TanimuraJCP92}
\begin{align}
  \frac{\partial }
{{\partial t}}W^{(n)} (p, q) ={}&  - ( { \mathcal{\hat L}_{QM}  + n\gamma } ) W^{(n)} (p, q) \nonumber\\
   &+  \frac{\partial }{{\partial p}}W^{(n+1)} (p, q)  \nonumber \\
&+ n\gamma \zeta  \!\left( {p + \frac{m}
{\beta }\frac{\partial }
{{\partial p}}} \right)W^{(n-1)}(p, q),
\label{eq:GMFPequation}
\end{align}
where $\zeta\equiv\eta/m$. These equations are a generalization of the Caldeira--Leggett quantum Fokker--Planck equations \cite{CaldeiraPhysica83,ChenWaxman1985} for  non-Markovian noise. \cite{TanimuraPRA91,TanimuraJCP92} The classical limit of the equations of motion can be obtained by taking the limit $\hbar \rightarrow 0$, which leads to the replacement of the Liouvillian through $\mathcal{\hat L}_\mathrm{QM} \to 
  \mathcal{\hat L}_\mathrm{cl} (p, q) \equiv    ({{p}/
{m})(\partial}/{\partial q})+f(q)({\partial}/
{\partial p})$, where $f(q)= {\partial U(q)}/{\partial q}$. The classical limit of Eq.~(\ref{eq:GMFPequation}) is the classical hierarchical Fokker--Planck  equations (CHFPE), which represent a generalization of the Kramers equation\cite{Kramers, Risken-Book} for non-Markovian noise. \cite{TanimuraPRA91,TanimuraJCP92} Because $\beta \hbar \gamma =0$ is always satisfied in the classical case, the validity of the CHFPE approach is not limited to the high-temperature regime in the classical case.

\subsection{Continued fraction expression and truncation of the hierarchical equations}
\begin{figure}[t]
\includegraphics[width=0.9\columnwidth]{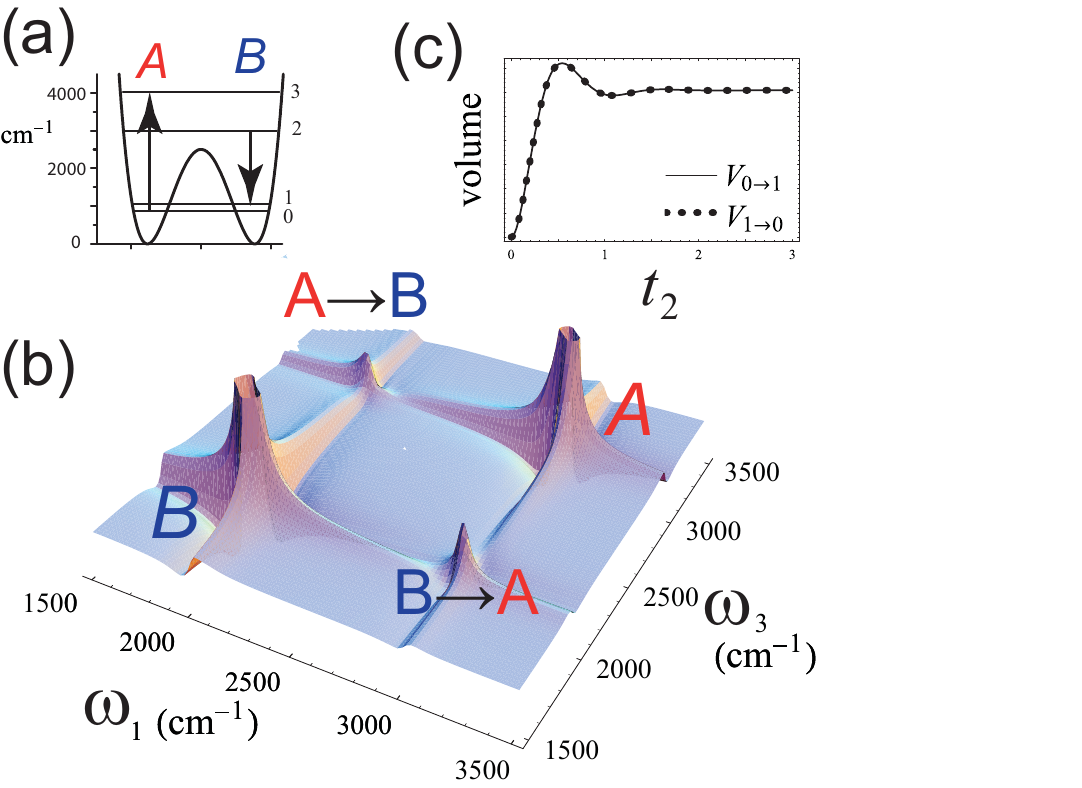}
\caption{\label{figPT2DIR:g}The ultrafast nonlinear laser spectrum is calculated from the nonlinear-response functions.
In a high-temperature case, the third-order optical response can be expressed analytically in terms of the HEOM resolvent in continued fraction form.\cite{Tanimura89B,Tanimura89C} Here, a chemical reaction process is discussed for a proton transfer system (a) coupled to a bath. Then, the two-dimensional (2D) vibrational spectrum $I(\omega_1, t_2, \omega_3)$ is calculated for different $t_2$ (see the Appendix). A snapshot of the 2D spectrum is presented in (b). The diagonal peaks arise respectively from the transitions $A$ and $B$ depicted in (a). The volume of the off-diagonal peaks has been shown to be proportional to the transition rate between the ground and the first excited state. By evaluating this volume as a function of $T_2$, we can evaluate the chemical reaction rate directly from the measurement (c). Reproduced from A. Ishizaki and Y. Tanimura, \href{https://doi.org/10.1143/JPSJ.58.1850} {J. Chem. Phys.} {\bf 123}, 14503 (2005), with the permission of AIP Publishing. }
\end{figure}

The hierarchy of equations introduced above continues to infinity and is ``formally exact,'' but these equations are not easy to solve numerically. If the system--bath coupling $\eta$ is small, the contribution of the elements of higher members of the hierarchy becomes smaller than that of the elements of the lower members; thus, we can safely disregard the deeper hierarchy. This is not the case, however, if the system--bath interaction is strong. In the high-temperature case, we can solve the HEOM analytically using the continued fraction expressions for the resolvent of a simple system.\cite{Tanimura89A,TanimuraPRA90}
Under the high-temperature condition of Eq.~\eqref{eq:HEOM}, the Laplace-transformed element $\hat{\rho}^{(0)}(t)$ with  factorized initial condition $\hat{\rho}^{(0)}(0)$ is expressed as ${\hat \rho}^{(0)}[s]={\hat G}_0[s]\hat{\rho}^{(0)}(0)$,  where the $n$th resolvent is defined in  recursive form as\cite{TanimuraJPSJ06,Tanimura89B} 
\begin{align}
  {\hat G}_n[s]=\cfrac{n+1}{s+(n+1)\gamma+i\hat{L}_A+\hat{\Phi}{\hat G}_{n+1}[s]{\hat \Theta_0}}.
\label{resolvent}
\end{align}
Nonlinear response functions can also be expressed as products of ${\hat G}_n[s]$ (see Fig.~\ref{figPT2DIR:g}), which allows us to evaluate the bathentanglement  analytically.\cite{Tanimura89B,Tanimura89C,IshizakiJCP05, Tanimura89D}  

For a complex system, most notably a system described in coordinate space or a system driven by a time-dependent external force, it is easier to solve the equations of motion with a finite number of the elements $K$ by adopting one of the following truncation schemes. The first  of these is the time-nonlocal scheme utilizing the asymptotic relation among the higher-order hierarchy members: As can be seen from Eq.~\eqref{resolvent}, we have ${\hat G}_K [s] \approx (K+1)/[s+(K+1)\gamma$] for $(K+1)\gamma \gg \omega_0$, where $\omega_0$ is the characteristic frequency of the system.\cite{TanimuraPRA91,TanimuraJCP92,TanimuraJPSJ06,Yan2015_truncation} The second  is the time-local scheme ignoring the higher-order equations of motion with  higher Matsubara frequency terms, while a renormalization operator is introduced to take into account the effects of the cutoff.\cite{IshizakiJPSJ05} 

The truncated HEOM can be evaluated with the desired accuracy by depicting the asymptotic behavior of the hierarchical elements for different $K$ and using this to determine whether or not there are sufficiently many members in the hierarchy. Essentially, the error introduced by the truncation becomes negligibly small when $K$ is sufficiently large.

Here, we illustrate their hybrid truncation scheme for the quantum Fokker--Planck equation (QFPE).\cite{Tanimura2015}
In this scheme, we set the number of Matsubara frequencies to be included in the calculation as $K$, which is chosen to satisfy $K \gg \omega_0 /\nu_1$. The upper limit for the number of hierarchy members for given $\gamma$ is chosen to be $K_{\gamma}\equiv \mathrm{int}( K \nu_1/\gamma )$ for $\nu_1>\gamma$ and $K_{\gamma}\equiv K$ for $\nu_1\le \gamma$. 
Then, for the case $\sum_{k = 1}^K {{j_k}} > K$, we truncate the hierarchical equations by replacing Eq.~\eqref{heom_wig} with
\begin{equation}
 \frac{\partial}{\partial t}{W}_\jk^{(n)}(p,q;t) = -( \mathcal{\hat L}_{QM} + \hat{\Xi'} ) W_\jk^{(n)}(p,q;t),
\label{wig_term}
\end{equation}
while, for the case $n=K_{\gamma}$ we employ
\begin{align}
 &\frac{\partial}{\partial t}{W}_{0, \ldots ,0}^{(K_{\gamma})}(p,q;t) =\nonumber\\
  &\qquad-( \mathcal{\hat L}_{QM}  + K_{\gamma}\gamma - \hat \Phi {{\hat {\bar \Theta} }_0}+ \hat{\Xi'}) W_{0, \ldots ,0}^{(K_{\gamma})}(p,q;t) \nonumber \\
	&\qquad  - K_{\gamma} \gamma {{\hat {\bar \Theta} }_0} 
W_{0, \ldots ,0}^{(K_{\gamma}-1)}(p,q;t),
\label{heom_wig_term}
\end{align}
where 
\[
\hat{\Xi'}\equiv \frac{m\zeta}{\beta}\left(\sum_{k=K+1}^{\infty} c_k \frac{2\gamma^2}{\gamma^2-\nu_k^2}  \right)\frac{\partial^2 }{\partial p^2}
\]
is the renormalization operator in the Wigner representation. The time-local scheme corresponds to Eq.~\eqref{wig_term} and the time-nonlocal scheme to  Eq.~\eqref{heom_wig_term}. We can then evaluate $W_{j_1,\dots,j_K}^{(n)}(p,q;t)$ through  numerical integration of the above equations. 
In the Markovian limit $\gamma \rightarrow \infty$,
which is taken after the high-temperature limit, yielding the condition  $\beta \hbar \gamma \ll 1$, we have the QFPE\cite{CaldeiraPhysica83,ChenWaxman1985}
\begin{align}
\frac{\partial }{{\partial t}}W^{(0)} (p, q; t) ={}& - \mathcal{\hat L}_{QM} W^{(0)} (p, q; t) \nonumber \\
	& + \zeta \frac{\partial}{\partial p}\!\left( {p + \frac{m} {\beta }\frac{\partial}{{\partial p}}} \right)
 W^{(0)} (p, q; t),
\label{eq:GWLLFokkerPlanck}
\end{align}
which is identical to the quantum master equation without the RWA.\cite{TanimuraJPSJ06}
Because we assume that the relation $\beta \hbar \gamma \ll 1$ is maintained, while taking the limit $\gamma \to \infty$,
this equation cannot be applied to low-temperature systems, in which quantum effects play a major role. As in the case of the master equation without the RWA, the positivity of the population distribution $P(q) = \int dp \,W(p,q;t)$ cannot be maintained if we apply this equation in the low-temperature case. To overcome this limitation, we must include  low-temperature correction terms to obtain the low-temperature-corrected QFPE (LT-QFPE).\cite{Ikeda2019Ohmic}

\subsection{Multistate quantum hierarchical Fokker--Planck equations}
\begin{figure}[t]
\includegraphics[width=\columnwidth]{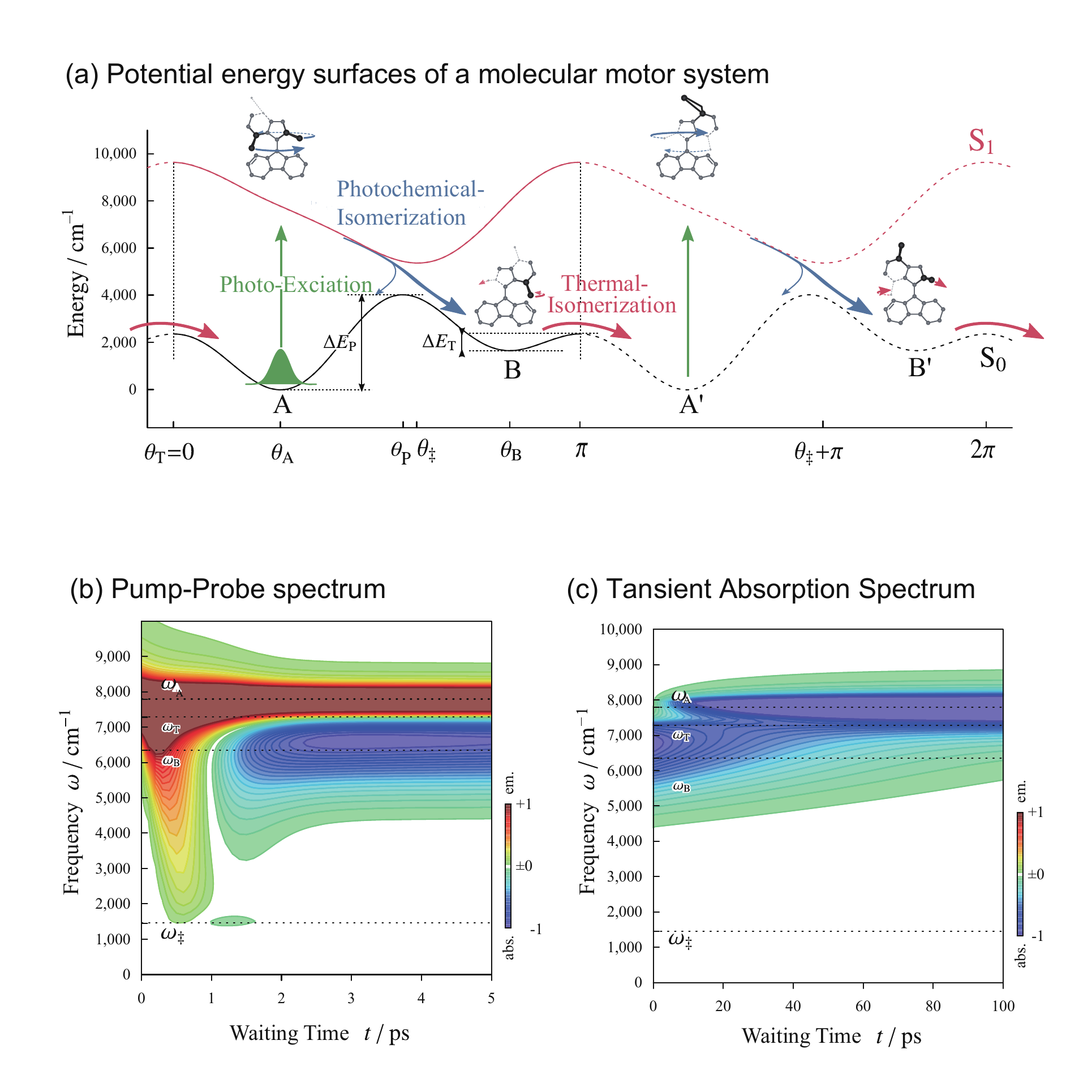}
\caption{\label{fig6MoterPPTA}The dynamics of a molecular motor system and its nonlinear optical response (see the Appendix) were investigated using the LT-MS-QSE approach.
 (a) The adiabatic Born--Oppenheimer potential energy surfaces (BOPESs) are plotted as functions of $\theta $. The black and red curves represent the ground ($\mathrm{S}_{0}$) and excited ($\mathrm{S}_{1}$) BOPESs, respectively. 
The thermal stable state and photoproduct state are labeled  ``A'' and ``B.'' Molecular structures of a typical two-step photodriven molecular rotary motor are also depicted. (b) Pump--probe spectrum (PPS) for the photoisomerization process from A to B. (c) Transient absorption spectrum (TAS) for the thermalization process from B to A. In both spectra, the red and blue areas represent emission and absorption, respectively. PPS is sensitive for detection of the photoisomerization process, while TAS is sensitive for detection of the thermalization process. Reproduced from T. Ikeda and Y. Tanimura, \href{https://doi.org/10.1063/1.4989537} {J. Chem. Phys.} {\bf 147}, 014102 (2017), with the permission of AIP Publishing.
}
\end{figure}
Up to this point, we have separately discussed two types of dynamics employing distinct approaches: one is a discrete energy-level system discussed from the viewpoint of the HEOM, and  the other is a molecular or atomic coordinate system discussed from the viewpoint of the QHFPE. As mentioned in Sec.~\ref{sec:System}, molecular motion and electronic excited states are coupled and play important roles simultaneously in many important physical processes. In the HEOM approach, this extension is straightforward and presented as the multistate QFPE, which can be applied to both optical and nonadiabatic transition problems, taking into account nonperturbative and non-Markovian system--bath interactions.
In this case, the Wigner distribution function is expanded in the electronic basis set as~\cite{MS-QFP94, MS-QFPChernyak96,Ikeda2017JCP,Ikeda2018CI}
\begin{align}
  {\bf W}(p, q)  = \sum_{j, k} | j \rangle W_{jk} (p, q) \langle k |,
  \label{eq:pjk}
\end{align}
where $W_{jk} (p, q)$ is the $jk$ element of the Wigner distribution function. 
We represent the Wigner functions for different electronic states in  matrix form as ${\bf W}(p, q) \equiv \{ W_{jk} (p, q) \}$.  
Similarly, the transition operator is written in  Wigner representation form as
\begin{align}
  X_{ij}^{\pm} (p, q; t)  = \pm i \int_{ - \infty }^\infty  dx 
\,e^{ ip x /\hbar  } 
U_{ij}\! \left( {q \mp \frac{x}{2}}; t \right). 
\label{eq:Xjk}
\end{align}
The quantum Liouvillian can also be expressed in  matrix form as\cite{MS-QFP94} 
\begin{align}
  &- \left\{ {\hat {{\bm{L}}}}_{QM} {\bf W}(p, q; t) \right\}_{jk}= \nonumber \\
  &\qquad -\frac{p}{m}\frac{\partial }{{\partial q}} W_{jk} (p, q;  t)  
  \nonumber \\ 
  &\qquad+ \frac{1}{\hbar } \int \frac{{dp'}}{{2\pi \hbar }}\sum \limits_m \left[ { X_{jm}^+ (p-p', q; t) W_{mk} (p', q; t) } \right. \nonumber \\
    &\qquad \qquad \qquad+ \left. {X_{mk}^{-} (p-p', q; t) W_{jm} (p', q; t)} \right].
  \label{eq:MSFP}
\end{align}
By using the above transformation, the hierarchy operators $\hat \Phi'$, ${\hat {\bar \Theta} }_0$ and $\hat{\Theta}_k'$, and $\hat{\Xi'}$ in Eq.~\eqref{heom_wig} are cast in  matrix form as $\hat {\bf \Phi'}$, ${\hat {\bar {\bf \Theta}} }_0$, $\hat{{\bf \Theta}}_k'$, and $\hat{{\bf \Xi'}}$ for the system--bath interaction defined as ${\bf V}(q)= \{ V_{jk}(q) \}$. In this way, we can obtain the multistate quantum hierarchical Fokker--Planck  equations (MS-QHFPE) for $W_{j_1,\dots,j_K}^{(n)} (p, q)$.\cite{TanimuraMaruyama97,MaruyamaTanimura98,Ikeda2017JCP,Ikeda2019JCP,Ikeda2018CI} 
Comparisons among the transient absorption spectrum calculated from the Ohmic MS-QHFPE with  low-temperature correction terms (LT-MS-QFPE), the Zusman equation, fewest switch surface hopping, and the Ehrenfest approach were made to demonstrate the reliability of the numerically  ``exact'' approach.\cite{Ikeda2019Ohmic} A numerical solution of the low-temperature multistate quantum Smoluchowski equation (LT-MS-QSE), which is the strong-damping limit of LT-MS-QFPE, is presented in Fig.~\ref{fig6MoterPPTA}. 

A quantum electron transport problem described by the spinless Anderson--Holstein model was also investigated in a similar manner.\cite{Thoss2018transerFPE}

\subsection{The  bathentanglement  states and the positivity of the HEOM}
\label{sec:Entangle}
\begin{figure}
\includegraphics[width=0.7\columnwidth]{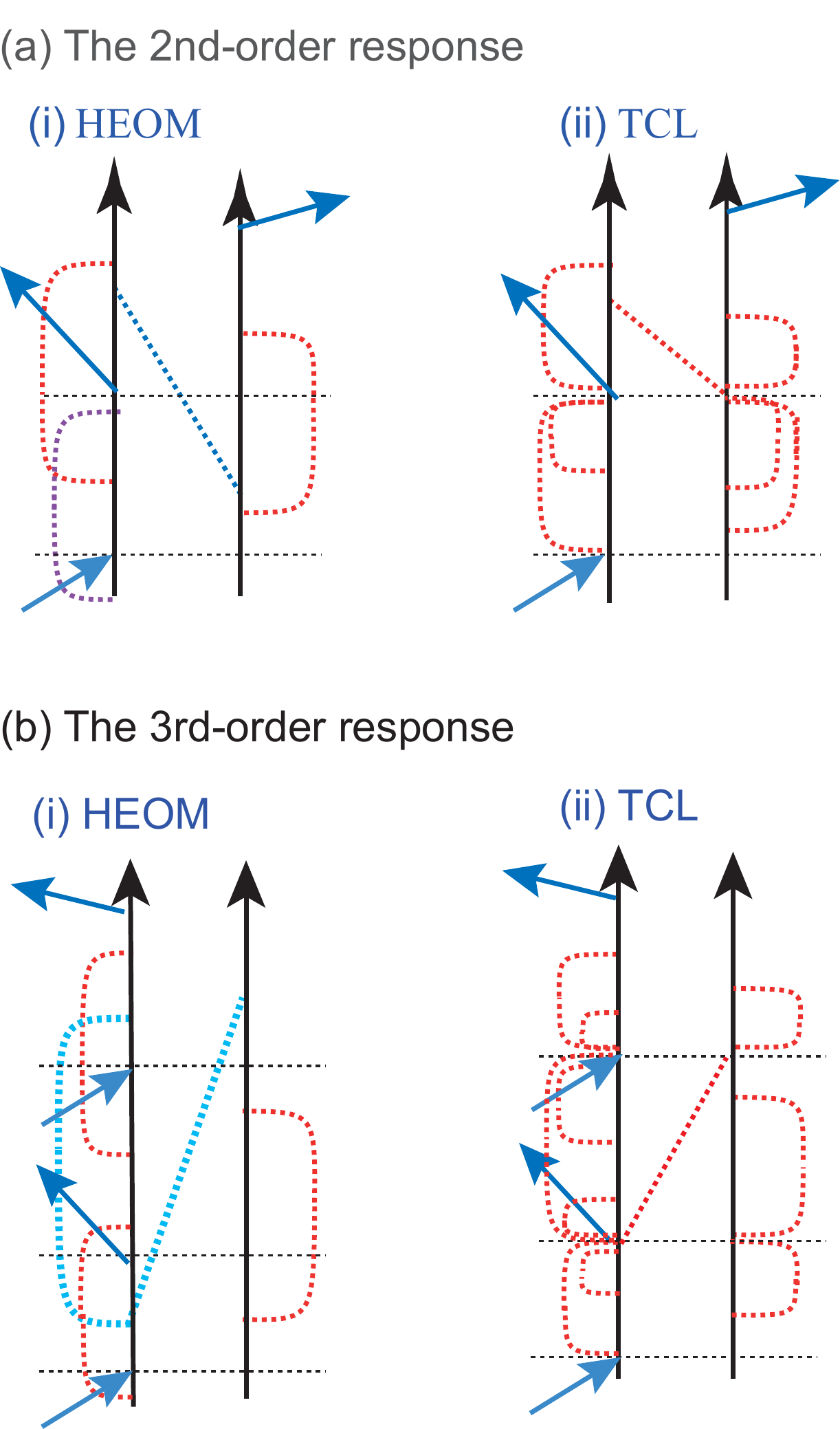}
\caption{Schematic diagrams involved in  (a) the second-order and (b) the third-order response functions, which are defined as the expectation values of the system under the two and three impulsive excitations illustrated by the blue arrows, respectively (see the Appendix).
The accurate and HEOM descriptions of the bathentanglement (system--bath entangled) states are illustrated in  (a-i) and (b-i), and the time-convolution-less (TCL) Redfield descriptions of these states are illustrated in  (a-ii) and (b-ii). Because of the factorization assumption, the TCL approach cannot take into account the  bathentanglement  at the time at which the external forces are applied to the system (the blue arrows with black dashed lines), whereas the HEOM can take these entangled states into account accurately by utilizing the hierarchy elements illustrated in Fig.~\ref{dsidedFyenman}. The simulation results corresponding to (a-i) and (a-ii) are presented in Figs.~\ref{fig:nonMarkov}(d-i) and (d-ii), and those corresponding to (b-i) and (b-ii) are presented in Figs.~\ref{LP}(a) and \ref{LP}(b), respectively. The blue dashed curves in (b-i) represent the  bathentanglement  states that contribute to the rephrasing echo signal in the 2D spectrum, as presented in Fig.~\ref{LP}(a).}
\label{fig:entangle}
\end{figure}

In the HEOM formalism, the information concerning the system--bath coherence (bathentanglement) is stored in the hierarchical elements, which allows us to simulate the quantum entangled dynamics between the system and bath (see Fig.~\ref{dsidedFyenman}). Thus, for example, the correlated initial equilibrium state can be set by running the HEOM program until all of the hierarchy elements reach a steady state and then use these elements as the initial state: the steady-state solution of the first hierarchy element $\hat \rho _{0,0, \dots ,0}^{(0)}(t=\infty)$ agrees with the correlated thermal equilibrium state defined by $\hat{\rho}_A^\mathrm{eq}= \mathrm{tr}_B\{\exp(-\beta \hat H_\mathrm{tot})\}$, while the other elements describe the  bathentanglement  states. Because the conventional Markovian and non-Markovian reduced equations of motion approaches eliminate the bath degrees of freedom completely, they cannot properly take into account the  bathentanglement  states, as illustrated in Fig.~\ref{fig:entangle}.\cite{IshizakiCP08, Shi2015CorretedInitial, Shi2012SBcorre,Tanimura2015}  Although the effects of  bathentanglement  are not easy to observe as long as we are working on linear response measurements, it is possible to study them experimentally by measuring the nonlinear response of the system (see the Appendix).

Because of their structure, the HEOM can treat any noise correlation even it becomes negative (as shown in Fig.~\ref{L_2Fig}), and the HEOM continue to satisfy the positivity condition, as demonstrated by various numerical simulations (see also Sec.~\ref{sec:nonMarkovinTests}). Owing to the complex structure of the HEOM, however,  analytical verification of their positivity has so far been limited to a simple case.\cite{Mintert2017} 

To obtain a numerically converged solution as the thermal equilibrium state, it is important to express the noise correlation function in terms of the decaying functions as $c_{j}\exp(-\zeta_{j} t)$ or $c_{j}\exp(-\zeta_{j} \pm i \omega_{j})$, where $c_{j}$, $\zeta_{j}$, and $\omega_{j}$ characterize the noise correlation for the ${j}$th hierarchical elements, to maintain the positivity of the HEOM formalism.\cite{TanimuraMukamelJPSJ94,TanakaJPSJ09,TanakaJCP10,YanBO12,Nori12, KramerFMO2DLorentz,TanimruaJCP12,Yan2011Drudebetter,Yan2017betterSpectraltreatmen,
KramerJPC2013BO2DEcho, Shi2014SpecralDecom, Ishizaki2020} Although the HEOM can be constructed for an arbitrary SDF in the form of a Fourier expansion\cite{TanimuraPRA90} or in terms of special functions, \cite{GHChen2012,Kleinekathoefer_Chevichev2016,Kleinekathoefer_Chevichev2019,WueHEOM2015,WueHEOM2017, WueHEOM2017JCP,WueHEOM2019JCP, Duan2020DuanCao} the HEOM that are derived from a finite set of  non-time-decaying functions do not converge in time and thus may not describe phenomena toward the thermal equilibrium state.

\subsection{Numerical techniques}
\label{sec:NumericalTech}
The HEOM are  simultaneous differential equations expressed in terms of the density matrix elements. Owing to the complex hierarchical structure, in particular in the low-temperature case, numerical integration of the HEOM is computationally intensive in terms of both memory and central processing unit (CPU), although, in many cases, there is no other way to obtain reliable results. Therefore, great efforts have been made to reduce the computational costs by improving the algorithmic and numerical techniques employed. An efficient approach is to suppress the number of elements in the hierarchy  by constructing the HEOM using Pad\'e\cite{YanPade10A,YanPade10B, YanPade11} or Fano \cite{YanFano19, YanFano20} spectral decompositions instead of Matsubara frequency decomposition. By introducing  rescaling factors, it is possible to reduce the number of elements in the hierarchy.\cite{ShiYan09rescal}
A numerical algorithm based on optimization of hierarchical basis sets\cite{YanOptimalBasis2015} and a tensor network has also been examined with the aim of reducing the number of calculations required.\cite{ShiJCP2018Tensor,BorrelliCP2018Tensor}

Like other reduced equation approaches, the HEOM are formulated in terms of the reduced density matrix, which requires $N \times N$ memory space for a system with  $N$ eigenstates. This makes the scalability of the system size very low, in particular when the system is described in  Wigner space. Therefore,  wave-function-based HEOM approaches whose scalabilities are similar to that of the Schr\"odinger equation have been developed \cite 
{Strunz2014,Strunz2015,Strunz2017A,Strunz2017B,Shi2016Sto,KeZhao2016,KeZhao2017, KeZhao2017B,KeZhao2018,KeZhao2019,WangKeZhao2019,Nakamura2018}
(see Sec.~\ref{sec:Evolution}C). In addition to the scalability of the memory, these approaches are advantageous because they allow us to employ the wide range of numerical techniques developed for the Schr\"odinger equation. 
 
Advances in computer technology have also led to  HEOM calculations becoming faster and larger. Thus,  computer codes for a graphic processing unit (GPU) and parallel computing through the Message Passing Interface (MPI) utilizing  distributed memory (DM) have been developed to treat large systems and various SDFs. Commonly used HEOM of this kind are the GPU-HEOM\cite{Kramer12GPU11,Kramer12NJP2Decho,Tsuchimoto2015} and DM-HEOM.\cite{Schuten12,KramerAlan14scalable,Kramer18DM-HEOM, Kramer18DM-HEOMB}
To develop an integral routine for the HEOM for a time step $\Delta t$, an easy and efficient way is to employ the math libraries developed for targeting computer architectures. For example, to utilize the GPU architecture, one can construct the large matrices for the entire HEOM reduced density operators and the Liouvillians, and then manipulate them using the math library suitable for these matrices.\cite{Tsuchimoto2015} The most efficient time-integration routine  currently available is the low-storage Runge--Kutta method,\cite{LSRK42017} which has been used to study  multistate nonadiabatic dynamics described in  two-dimensional configuration space.\cite{Ikeda2018CI}

\section{Numerically ``exact'' tests for non-Markovian and nonperturbative dynamics}
\label{sec:nonMarkovinTests}
\begin{figure}[t!]
\includegraphics[width=\columnwidth]{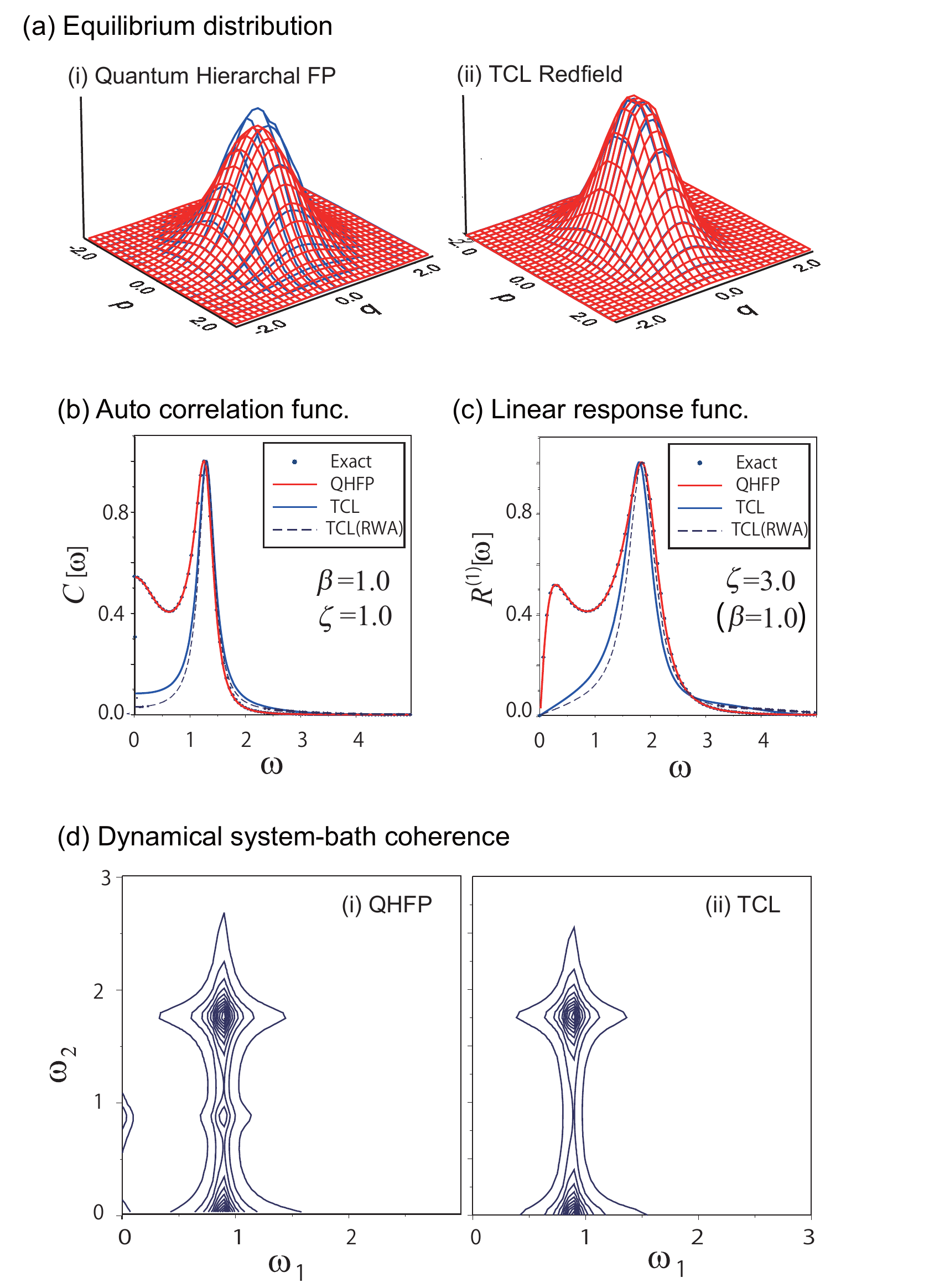} 
\caption{Numerically ``exact'' tests for the description of (a) the  bathentanglement  thermal equilibrium state, (b) fluctuations, (c) dissipation, and (d) dynamical  bathentanglement  calculated for a Brownian oscillator system with  frequency $\omega_0=1.0$.\cite{Tanimura2015} The values of the bath parameters  for the Drude SDF were chosen as $\gamma=1.0$, $\zeta=1$, and $\beta\hbar=3.0$, unless otherwise indicated. (a) Factorized initial conditions (blue curves) and steady-state solutions (red curves), calculated from (i) the QHFPE and (ii) the TCL Redfield equation. (b) Autocorrelation function of the Brownian oscillator system for $\beta\hbar=1.0$, and (c) the linear response functions for $\zeta=3.0$. Here, the dotted, red, blue, and blue-dashed curves represent the results obtained from the analytical expression, the QHFPE, the TCL Redfield equation, and the TCL Redfield equation with the RWA, respectively. (d)  Second-order response function of the Brownian oscillator system, $R^{(2)}_\mathrm{TTR}[\omega_1, \omega_2]$.\cite{TaniMuka2D, Ikeda2015} The results obtained from the QHFPE approach (i) almost overlap with the analytical result, while the results from the TCL Redfield approach (ii) miss some peaks owing to the factorized nature of the TCL description, as illustrated in Fig.~\ref{fig:entangle}(a).  \href{https://doi.org/10.1063/1.4916647} {J. Chem. Phys.} {\bf 142}, 144110 (2015); licensed under a Creative Commons Attribution (CC BY) license.
}
\label{fig:nonMarkov}
\end{figure}

The capabilities of the reduced dynamics theory can be verified through nonperturbative and non-Markovian tests on the basis of  analytical solutions for the Brownian oscillator model with arbitrary SDF, $J(\omega)$.\cite{Tanimura2015} These tests can be applied to the system described in the coordinate representation, $\hat{H}_A = \hat{p}^2/{2m} + m\omega_0^2\hat{q}^2/2$, or to the system in the energy eigenstate representation, $\hat{H}_A= \hbar \omega_0({\hat a}^+{ \hat a}^-+1/2)$, where ${\hat a}^{\pm}$ are the creation--annihilation operators for the eigenstates.  The tests are based on the solutions for (a) the steady-state distribution, (b) the symmetric autocorrelation function $C(t)\equiv \langle \hat q(t) \hat q+ \hat q \hat q(t) \rangle/2$, (c) the linear response function $R^{(1)}(t)\equiv i\langle [\hat q(t), \hat q] \rangle/\hbar$, and (d) the nonlinear response function $R_\mathrm{TTR}^{(2)}(t_2,t_1)=-\langle[[\hat q^2 (t_1+t_2), \hat q(t_1)], \hat q] \rangle/{\hbar^2}$,\cite{TaniMuka2D,Ikeda2015} which is the observable of  2D terahertz Raman spectroscopy,\cite{Phamm2DTHzRaman} (the simulation method is explained in the Appendix).
 These are tests of the ability of the theory to account for (a) the  bathentanglement  thermal equilibrium state, (b) fluctuations, (c) dissipation, and (d)  dynamical  bathentanglement. Test (d) is particularly important if we wish to study dynamics under time-dependent external forces.\cite{TanimuraMukamelJPSJ94,TanimuraMaruyama97,MaruyamaTanimura98,Yan2009extfielf} 
Here, we illustrate these results for the Drude case [Eq.~\eqref{JDrude}] calculated from analytically exact expressions for the Brownian oscillator system,\cite{GrabertZPhys84, GrabertPR88, Weiss08} the QHFPE,\cite{Tanimura2015} and the TCL Redfield equation.\cite{TCLshibata77,TCLshibata79,TCLshibata10} 
For  cases (b)--(d), we plot the Fourier-transformed results denoted by $C[\omega]$, $R^{(1)}[\omega]$, and $R^{(2)}_\mathrm{TTR}[\omega_1, \omega_2]$. 

Figure~\ref{fig:nonMarkov}(a) illustrates the factorized initial conditions (blue curves) and steady-state solutions (red curves) calculated from (i) the QHFPE and (ii) the TCL Redfield equation. After integrating the QHFPE and the TCL Redfield equation for a sufficiently long time, the distribution reaches a steady state. In the case of the QHFPE, the obtained steady state is identical within numerical error to the thermal equilibrium state obtained from the analytical expression, whereas those from the TCL Redfield equation are similar to the original factorized initial state: the difference between the two distributions arises from  ``static bathentanglement'' and represents the nonfactorized effect of the thermal equilibrium state. 

Figure~\ref{fig:nonMarkov}(b) depicts the autocorrelation (symmetric correlation) functions in the frequency domain. The dotted, red, blue, and blue-dashed curves represent the results obtained from the analytical expression, the QHFPE, the TCL Redfield equation, and the TCL Redfield equation with the RWA, respectively. 

Figure~\ref{fig:nonMarkov}(c) shows the linear response function $R^{(1)}[\omega]$ for  strong system--bath coupling, $\zeta=3.0$. This function is temperature-independent in the harmonic case. While the QHFPE results (red curves) coincide with the exact results (black dots), the TCL Redfield results without the RWA (blue curves) and with the RWA (blue dashed curves) are close only near the maximum peak, regardless of temperature. The low-frequency parts of the spectra arise from the slow dynamics of the reduced system near the thermal equilibrium state, and the discrepancy between the TCL results and the exact results arises from the slow equilibration process.

Figure~\ref{fig:nonMarkov}(d) shows the second-order response function $R^{(2)}_\mathrm{TTR}[\omega_1, \omega_2]$ corresponding to the intermediate coupling case considered in Fig.~\ref{fig:nonMarkov}(b). The results here were obtained from (i) the QHFPE approach and (ii) the TCL Redfield approach without the RWA. The QHFPE results overlap with those from the analytical expression. Owing to  ``dynamical bathentanglement,'' we observe  peaks near $(\omega_1,\omega_2)=(0,1)$ and $(1,1)$, whereas the conventional reduced equation of motion approaches cannot take system--bath coherence into account  owing to their use of a factorized description of the bath state, as shown in Fig.~\ref{fig:entangle}(a).

As illustrated in this section, we have been able to obtain very accurate results from the HEOM approach, as judged by  tests (a)--(d). By contrast, we have found that the TCL Redfield equation has limited applicability when judged in a similar way.  

\section{HEOM: Time-Evolution Equations}
\label{sec:Evolution}
\subsection{HEOM for arbitrary spectral distribution functions}
To extend the applicability of the HEOM in a numerical ``exact'' manner, various HEOM have been developed. A straightforward extension of the HEOM approach is to consider a more complicated SDF to describe the environmental effects of realistic chemical and biochemical systems. When the noise correlation function is expressed in terms of  damped oscillators, which can be done  for the Brownian,\cite{TanimuraMukamelJPSJ94,TanakaJPSJ09,TanakaJCP10, YanBO12} Ohmic,\cite{Ikeda2019Ohmic} Lorentz,\cite{Nori12} super-Ohmic,\cite{Shi2014SpecralDecom} Drude--Lorentz\cite{KramerFMO2DLorentz} SDFs (and combinations of these; see Refs.~\onlinecite{TanimruaJCP12, KramerJPC2013BO2DEcho}), we can obtain the reduced equations of motion in the hierarchical structure.
Alternatively, we can extend the HEOM for arbitrary SDFs using  Fourier,\cite{TanimuraPRA91} Gauss--Legendre, or Chebyshev--quadrature spectral decompositions (eHEOM).\cite{GHChen2012,Kleinekathoefer_Chevichev2016,Kleinekathoefer_Chevichev2019,WueHEOM2015,WueHEOM2017, WueHEOM2017JCP, WueHEOM2019JCP, Duan2020DuanCao} This approach allowed the investigation of a spin-boson system coupled to a bath with sub-Ohmic SDF at  zero temperature.\cite{WueHEOM2017, WueHEOM2017JCP, WueHEOM2019JCP} An extension of the HEOM for a combination of exponential/non-exponential bath correlation functions has also been presented.\cite{Ikeda2020}

\subsection{Stochastic HEOM}
\label{subsec:StochasticHEOM}
The difficulty of solving the HEOM arises from the fluctuation terms described by $L_2(t)$, which give a negative contribution at low temperature (Fig.~\ref{L_2Fig}). In the classical generalized Langevin formalism, dissipation is described by a damping kernel, while  fluctuations are described by  noise, whose correlation function is related to the damping kernel through the fluctuation--dissipation theorem. Although the trajectories of a particle under  negatively correlated noise cannot be evaluated using the Langevin formalism, we can evaluate the distribution function of the trajectories by extending the HEOM formalism. \cite{TanimuraJPSJ06} Thus,  various stochastic equations of motion have been derived in which a hierarchical structure is employed for the dissipation part, while the fluctuation part is treated as noise.\cite{CaoStochastic2013,CaoStochastic2018A,CaoStochastic2018B,Shao2004} 
The efficiency of this approach relies on its treatment of the noise correlation. A methodology that treats the exponentially decaying part of the noise using a hierarchical formulation in  the same manner as the dissipation has been developed (the stochastic HEOM). Because this approach does not depend on a hierarchy of Matsubara frequencies or a Pad\'e decomposition, the memory requirement of computations is dramatically reduced in comparison with that of the regular HEOM, in particular for a system interacting with multiple heat baths. An extension of the stochastic approach to non-Gaussian noise that includes  noise from a spin bath has also been suggested.\cite{CaoStochastic2018B}  However, in the stochastic approach, although the memory requirement  is reduced,  sampling of the trajectories become difficult for the lower-temperature case or for longer-time simulations, in particular when the system is driven by a  time-dependent external force. Thus, under such conditions, the accuracy of calculations becomes lower than that of the regular HEOM.

\subsection{Wave-function-based HEOM}
\label{subsec:WfuncHEOM}
\begin{figure}
\includegraphics[width=0.4\columnwidth]{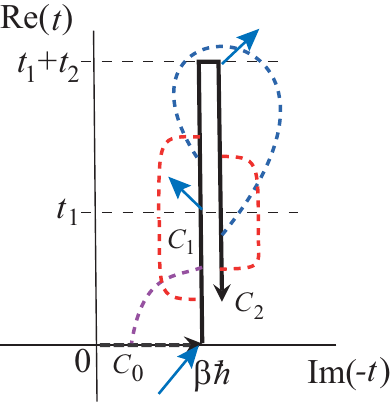}
\caption{\label{CP}The complex counter path for the hierarchical Schr\"odinger equations of motion (HSEOM) \cite{Nakamura2018} depicted with the system--bath interaction for the Liouville path of the second-order response function presented in Fig.~\ref{fig:entangle}(a-i).
}
\end{figure}

As shown in the derivation of the influence functional with a correlated initial condition, the Feynman--Vernon influence functional can also be defined on the basis of the wave function using complex time counter integrals. \cite{Tanimura2014}
This indicates that we can derive  HEOM for a wave function: such approaches   include the stochastic hierarchy of pure states (HOPS),\cite{Strunz2014,Strunz2015,Strunz2017A,Strunz2017B} the stochastic Schr\"odinger equation (SSE),\cite{Shi2016Sto}  the hierarchy of stochastic Schr\"odinger equations (HSSE), \cite{KeZhao2016,KeZhao2017, KeZhao2017B,KeZhao2018,KeZhao2019,WangKeZhao2019} and the hierarchical Schr\"odinger equations of motion (HSEOM),\cite{Nakamura2018} all of whose scalabilities are similar to that of the Schr\"odinger equation. These equations are advantageous not only because the scale of the  memory required becomes $N$ for an $N$-level system, but also because various numerical techniques developed for the Schr\"odinger equation can be employed for configuration space or energy eigenstate representations or a mixture of these.  Technically, these wave-function-based approaches are not regarded as  equation of motion approaches, because they are not defined by the time $t$ but by the complex time counter path $0 \rightarrow -i\beta\hbar \rightarrow t-i\beta\hbar \rightarrow -i\beta\hbar $ (see Fig.~\ref{CP}). Thus, whereas  in the conventional HEOM approach, the  HEOM density operators $\hat{\rho}_{j_1,\dots,j_K}^{(n)}(t+\Delta t)$ are computed from $\hat{\rho}_{j_1,\dots,j_K}^{(n)}(t)$, in the wave-function-based HEOM approach, we must  repeat the full integration for $t$ along $ -i\beta\hbar \rightarrow t+\Delta t -i\beta\hbar \rightarrow -i\beta\hbar $ for different $\Delta t $. Moreover, because the convergence of the trajectories becomes slow in the stochastic approaches and because the required Chebyshev function set becomes larger for longer simulations in the HSEOM approach, the wave-function-based approach is not suitable for studying a system with slow relaxation or a system subjected to a slowly varying time-dependent external force. This indicates that the regular HEOM are the method of choice to study long-term dynamics if the system is not too large, while the wave-function-based HEOM give access to  short- and intermediate-time dynamics over a wider range of temperatures for a large system.

\subsection{HEOM for a grand canonical electron bath}
\begin{figure}[t]
\includegraphics[width=\columnwidth]{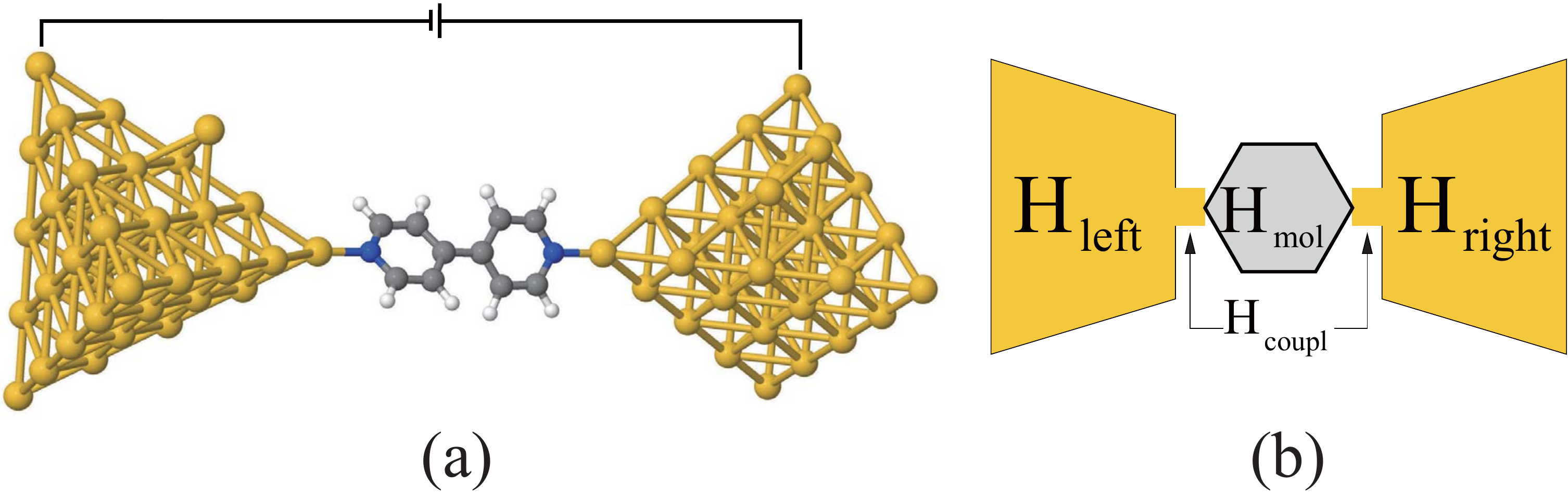}
\caption{Atomistic (a) and schematic (b) models of a molecular junction, showing the left and right electrodes (leads) and the molecular bridge. 
Reproduced from M. Thoss and F. Evers, \href{https://doi.org/10.1063/1.5003306} {J. Chem. Phys.} {\bf 148}, 030901 (2018), with the permission of AIP Publishing.
} \label{E-Vcoupling}
\end{figure}
\label{sec:fermionicHEOM}

Although the HEOM were originally obtained by considering a harmonic heat bath in a canonical ensemble,  Yan and his collaborators\cite{Yan2007HGrandCanonical} showed that the HEOM could also be  derived from  Gaussian-fermionic and Gaussian-bosonic bath models in a grand canonical ensemble by introducing the chemical potential $\mu$ to serve as a second thermodynamic variable in addition to the temperature. The HEOM for an electron bath is appropriate for studying charge transport problems with electron--electron interactions, most notably a quantum dot and a molecule  coupled to two electrodes and representing a continuum of noninteracting electronic states \cite{JinYanFermi08} (see Fig.~\ref{E-Vcoupling}).

We consider a Hamiltonian $\hat H_A$ described in terms of the creation and annihilation operators of the $j$th electronic state $\hat d_{{j;\sigma}}$ and $\hat d_{{j;\sigma}}^{\dagger}$, where $\sigma$ represents the spin state. The electrode bath is expressed as 
\begin{align}
\hat H_{B} = \sum_{k;\sigma} {\hat c}_{k;{\sigma}} ^{\dagger}{\hat c}_{k;{\sigma'}}, 
\end{align}
where ${\hat c}_{k;{\sigma'}}$ and ${\hat c}_{k;{\sigma'}} ^{\dagger}$ are the creation and annihilation operators of an electron in an electrode with  wavevector ${\bf k}$ and spin $\sigma'$. The system--electrode interaction is expressed as  
\begin{align}
\hat H_{I} = \sum_{f;\sigma}\sum_{k;\sigma'} V_{k;{\sigma'}}({\hat c}_{k;{\sigma'}} ^{\dagger} {\hat d}_{f;{\sigma}}+{\hat c}_{k;{\sigma'}}{\hat d}_{f;{\sigma}}^{\dagger}), 
\end{align}
where $f$ represents the  site of the system that is coupled to an electrode, and $V_{k;{\sigma'}}^{\alpha}$ describe the tunneling rates for the left and right electrodes, with $\alpha=L$ and $R$, respectively.  The electrode system is then characterized by the tunneling efficiency function 
\begin{align}
\Gamma_{\alpha}[\epsilon] = 2 \pi  \sum_{j;\sigma'}  (V_{j;{\sigma'}}^{\alpha})^2 \delta(\epsilon -\epsilon_{j;\sigma'} ),
\end{align}
which plays a similar role to the harmonic SDF.  

For the sake of conciseness, here we consider a spinless electronic bath where the system-electrode coupling and the bandwidths of the electrodes are assumed to be symmetric with respect to both sides of the lead. Then, the effective tunneling distribution is assumed to have a simple Drude form
\begin{align}
\Gamma_{\alpha}[\omega]=\frac{2 \pi \zeta_{\alpha} \gamma}{(\omega - \mu_{\alpha})^2+\gamma^2},
\end{align}
where $\zeta_{\alpha}$ is the coupling strength and $\mu_{\alpha}$ is the chemical potential for $\alpha=L$ and $R$.  This allows us to express the electronic bath relaxation function in the form 
\begin{align}
\Gamma_{\alpha}^s (t)=\sum_{m=0}^n \eta_m^{\alpha,s} e^{-\gamma_m^{\alpha,s} t},
\end{align}
with 
 \[
 \gamma_0^{\alpha,s} = \gamma-si\mu_{\alpha}, \qquad \eta_0^{\alpha,s} =\frac{\pi\zeta_{\alpha}\gamma}{1+e^{i\beta \gamma}},
 \]
\begin{align*}
\gamma_m^{\alpha, s}&= \beta^{-1}(2 m-1) \pi -si \mu_{\alpha}, 
\end{align*}
and
\begin{align*}
 \eta_m^{\alpha,s} =\frac{-i2 \pi\zeta_{\alpha}\gamma^2}{ \beta [(s\mu_{\alpha}-i\gamma_m^{\alpha,s})^2+\gamma^2]}
\end{align*}
for $m\ge1$. The index $s \in \{+,-\}$ corresponds to  creation and annihilation of electrons, i.e., $\hat d^{+}_{j;\sigma}=\hat d^{\dagger}_{j;\sigma}$ and $\hat d^{-}_{j;\sigma}=\hat d_{j;\sigma}$. Thus, the HEOM for a system coupled to the electrode baths can be expressed as\cite{JinYanFermi08,Reichman2013qdot, Thossperspective2018}
\begin{widetext}
	\begin{align}
	\frac{\partial}{\partial t}
	\hat{\rho}_{\bm{a}_1,\dots,  \bm{a}_n}(t)
	={}&-\left(
	i \hat{ L}_A+ 
	\sum_{k=1}^n \gamma_{m_k}^{\alpha_k,s_k}
	\right)
	\hat{\rho}_{{\bm{a}_1,\dots,  \bm{a}_n}}(t) \notag\\
	&-i \sum_{ \bm{a} } \left[ 
	\hat d_{j;\sigma}^{\bar{s}}  \hat{\rho}_{\bm{a}_1,\dots, \bm{a}_n,  \bm{a}}(t)-
	(-)^{n}   \hat{\rho}_{\bm{a}_1,\dots, \bm{a}_n, \,\bm{a}}(t) \hat d_{j;\sigma}^{\bar{s}} \right]
	\notag \\ 
	&-i  \sum_{k=1}^n (-)^{n-k} \left[  \eta_{m_k}^{\alpha_k,s_k}
	\hat d_{j_k;\sigma_k}^{s_k} \hat{\rho}_{\bm{a}_1,\dots,  \bm{a}_{k-1},  \bm{a}_{k+1}, \dots, \,\bm{a}_n}(t) 
 +
(-)^{n} (\eta_{m_k}^{\alpha_k,\bar{s}_k})^* \hat{\rho}_{\bm{a}_1, \dots,  \bm{a}_{k-1}, \, \bm{a}_{k+1}, \dots, \, \bm{a}_n}(t)
	\hat d_{j_k;\sigma_k}^{s_k}
	\right],
	\label{eq:FermiHEOM}
	\end{align}
	\end{widetext}
where $\bm{a}_k =\{j_k,\alpha_k,\sigma_k, m_k, s_k\}$ and $\sigma_k$ is the index of the system side of the spin in the $j_k$th electronic state, $m_k$ is the index for the Matsubara frequencies, and $s_k$ can be $+$ or $-$ with $\bar{s}_k=-s_k$. 	
 Because the higher-order cumulants vanish in the treatment of a Gaussian-electron source, it is not necessary to utilize the influence functional.\cite{ChernyakYujin2019}
At first glance, the above HEOM appear to be similar to the conventional HEOM, but their structure is more complex than that of the regular HEOM, because the number of elements in this kind of hierarchy quickly increases as the  system size increases, analogously to the HEOM derived on the basis of  Fourier decomposition.\cite{TanimuraPRA90} Extensions to take  account of a realistic band structure\cite{Thoss2019Chargetransport,Thoss2018transerBandstructure} and a truncation scheme specific for a fermionic bath have been investigated.\cite{YanFermi18truncation} A computer code for a fermionic bath,  HEOM-Quick, has been developed and distributed.\cite{Yan2016WIREs}
The electrode heat bath considered here is a noncorrelated electron environment based on a Gaussian single-particle picture, where  fermionic fluctuation--dissipation at second order can be applied. This condition is convenient when a density functional treatment of the system is employed. Thus, the real-time evolution of the reduced single-electron density matrix at the tight-binding level has been studied by combining time-dependent density functional theory with HEOM theory (TDDFT-HEOM).\cite{DFTHEOMYan2010,DFTHEOMChen2012}

An extension to include electronic--vibrational coupling has also been presented.\cite{Yan2016e-vApprox} The expression for the HEOM in this case was derived using a small polaron transformation to take account of strong electronic--vibrational coupling.\cite{Thoss2016e-vCouple,Thoss2018e-vCouple,Thoss2020full_conunt}

\subsection{HEOM for different system--bath models}
The HEOM approach has also been applied to the case of solid state materials described by a deformation potential\cite{TanimuraPRA90} and the Holstein Hamiltonian. \cite{LipenHolns2015,ReichmanHolns2019} In the Holstein case, because we can study only a small system and the number of phonon modes associated with the system site is finite,  special treatment is necessary to maintain the stability of the equations.\cite{ReichmanHolns2019} 

The HEOM for a molecular rotor system in which the rotational symmetry of the system is maintained have also been derived.\cite{Lipeng2019,Iwamoto2019} This extension is important to account for the rotational bands of linear and nonlinear spectra.

\subsection{Phenomenological and approximate approaches}
Although it is not possible to construct a reduced equation of motion in the HEOM structure for various system--bath Hamiltonians through a procedure of self-induction, one can still adopt the HEOM structure in a phenomenological approach.\cite{Yan16Projection_approximation,Yan2009whiteclassicalFP}  For  example, if the heat bath is not harmonic or not Gaussian owing to  anharmonicity of the bath oscillators or to nonlinearity of the system--bath interactions, we have to include  hierarchical elements with  odd orders of  noise correlation functions, most typically $\langle [ {\hat x}_j^2 (t_1), {\hat x}_j (0)] \rangle$ and $\langle [{\hat x}_j (t_2),[ {\hat x}_j (t_1), {\hat x}_j (0)]] \rangle$ for the bath coordinate ${\hat x}_j$ for various  time sequences $t_i$. The contributions from even orders of noise correlation functions, for example, $\langle [{\hat x}_j (t_3), [{\hat x}_j (t_2),[ {\hat x}_j (t_1), {\hat x}_j (0)]]]\rangle$, must be taken into account, because the higher-order cumulants do not vanish for a non-Gaussian bath. Accordingly, we have to introduce an SDF for these higher-order noise correlation functions, as  has been investigated in  molecular dynamics simulations.\cite{Okazaki1999} Complex profiles of these correlation functions have been investigated as the observables of  multidimensional vibrational spectroscopy.\cite{TaniMuka2D,OkuTani2D} For  electron transport in nanosystems, non-Gaussian dynamics were investigated using the HEOM approach in the framework of full counting statistics.\cite{Thoss2020full_conunt}

It should be noted that in a non-Gaussian case, either a fermionic or bosonic case, the regular fluctuation--dissipation theorem cannot be applied, because the bath response is highly nonlinear. 
Nevertheless, assuming that the non-Gaussian effects are minor, we can still adopt the HEOM structure as the starting point to simulate the system dynamics.\cite{DEOM2014,DEOM2018} Non-Gaussian effects can then be included as non-Gaussian corrections in the HEOM formalism.\cite{nonGaussPRL}

As an approximate approach, the HEOM have also been utilized to construct a time-dependent kernel of an open quantum dynamics equation for a charge carrier transfer process with Holstein--Peierls interactions. \cite{Shi2019ChargeCarrier}

\subsection{Imaginary-time HEOM}
Although we can obtain the correct (entangled) thermal equilibrium state of the reduced system, as the steady-state solution of the HEOM, it is easier to solve the imaginary-time HEOM by integrating over the inverse temperature.\cite{Tanimura2014,Tanimura2015}
Because the structures of the imaginary-time HEOM and the wave-function-based HEOM are similar, the cost of numerical integration is much cheaper than that of the regular HEOM. Moreover, because the solution of the imaginary-time HEOM is the partition function rather than the distribution function, we can evaluate thermodynamic variables, most importantly the Helmholtz free energy and the Boltzmann entropy, as functions of  temperature.\cite{Tanimura2014,Tanimura2015} 

If we wish to obtain a steady-state solution under a periodic external force, however, we must use the real-time HEOM. Therefore, a method has been developed to obtain  steady-state solutions of the HEOM very efficiently. \cite{ZhengYanFermi08,Yan17steadySolver}

\section{Applications}
\label{sec:Applications}

\subsection{Proton, electron, and exciton transfer problems}

Many chemical physics and biochemical physics problems involve environments that are complex and strongly coupled to a molecular system at finite temperature.
Moreover, the non-Markovian effects of environments arising from  intermolecular and intramolecular interactions play essential roles. Therefore, a great deal of effort has been dedicated to studying the problems of quantum as well as classical dynamics with nonperturbative and non-Markovian system--environment interactions from the HEOM approach. Commonly studied problems of this kind are chemical reactions,\cite{TanimuraPRA91,TanimuraJCP92, IshizakiJCP05} proton transfer, \cite{Shi2011PT,Jianji2020} electron transfer,\cite{TanakaJPSJ09,TanakaJCP10, Shi2009ET, TanimruaJCP12} electron-coupled proton transfer,\cite{Shi2017PCET} exciton-coupled electron transfer, \cite{Sakamoto2017} charge separation,\cite{Shi2018Chargeseparation} exciton--hole separation,\cite{KatoIshi2018} and nonadiabatic transitions in photochemical processes,\cite{TanimuraMaruyama97, MaruyamaTanimura98,Ikeda2017JCP,Ikeda2019JCP} including those involving conical intersection\cite{Lipen2016CI, Duan2016CI,Duan2017CI,Desouter-Lecomte19CI} and exciton transfer.\cite{Ishizaki09,Ishizaki2010A,Ishizaki2010B, Shi11FMO2DES,Ishizaki2011,Ishizaki2013, Yan2011_ChineseJ,Yan2010_2DES,Yan2013_2DES,Aspru11, CaoStochastic2013,Shi2013,OliverTonu2015exciton,Mancal20185exciton,Fujihashi2015,Fujihashi2017,Fujihashi2019,OliverTonu2015exciton,Mancal20185exciton,DijkstraNJP12exciton, Schuten09,Schuten11,DijkstraNJP10DNA,DijkstraNJP12exciton, Dijkstra2015fifth,Dijkstra2019_2DBOS,DijkstraProkorenko2017, SaitoFMO2019, YaoYaoJCP14,CaoPRB2012, YanMo2020} In particular, the HEOM approach has been used to simulate the exciton transfer process of the photosynthesis antenna system with the aim of investigating how natural systems can realize such highly efficient yields, presumably by manipulating quantum mechanical processes. This is because the HEOM approach can be applied seamlessly across a wide coupling range from the weak Redfield regime to strong F\"orster exciton--environment coupling under various non-Markovian conditions,\cite{Ishizaki2009B} and because it allows  calculation of multidimensional electronic spectra, which are the observables of  experiments of this kind.\cite{Kramer12NJP2Decho} 

The numerical costs of solving the HEOM for photosynthesis problems are high, because each exciton or electron site is independently coupled with its own heat bath. For  exciton transfer problems, on the other hand, because the systems are in the high-temperature regime, where the quantum nature of thermal noise plays a minor role, we can integrate the HEOM by ignoring the low-temperature correction terms. Thus, although the Fenna--Matthews--Olsen (FMO) system has often been used as a benchmark test, it is not suitable to verify the description of  open quantum dynamics, (however, it is convenient for testing the scalability of calculations, in particular for a realistic SDF). As a test of open quantum dynamics, the non-Markovian tests presented in Sec.~\ref{sec:nonMarkovinTests} are more suitable.

The ability of the HEOM to deal with time-dependent external forces is ideal for investigating the possibility of Floquet engineering for  exciton transfer processes\cite{IshizakiPhuc2018} and electron transfer,\cite{IshizakiPhuc2019A} in which the system Hamiltonian involves a periodic external force. On the basis of the HEOM approach, the effects of nondissipative local phonon modes have been investigated. Examples of this kind are  systems described by the Holstein model\cite{LipenHolns2015,ReichmanHolns2019} and the Holstein--Tavis--Cummings (HTC) model.\cite{IshizakiPhuc2019B} 

\subsection{Nonlinear and multidimensional spectroscopies}
\label{subsec:2Dspectra}
\begin{figure}[t]
\includegraphics[width=0.95\columnwidth]{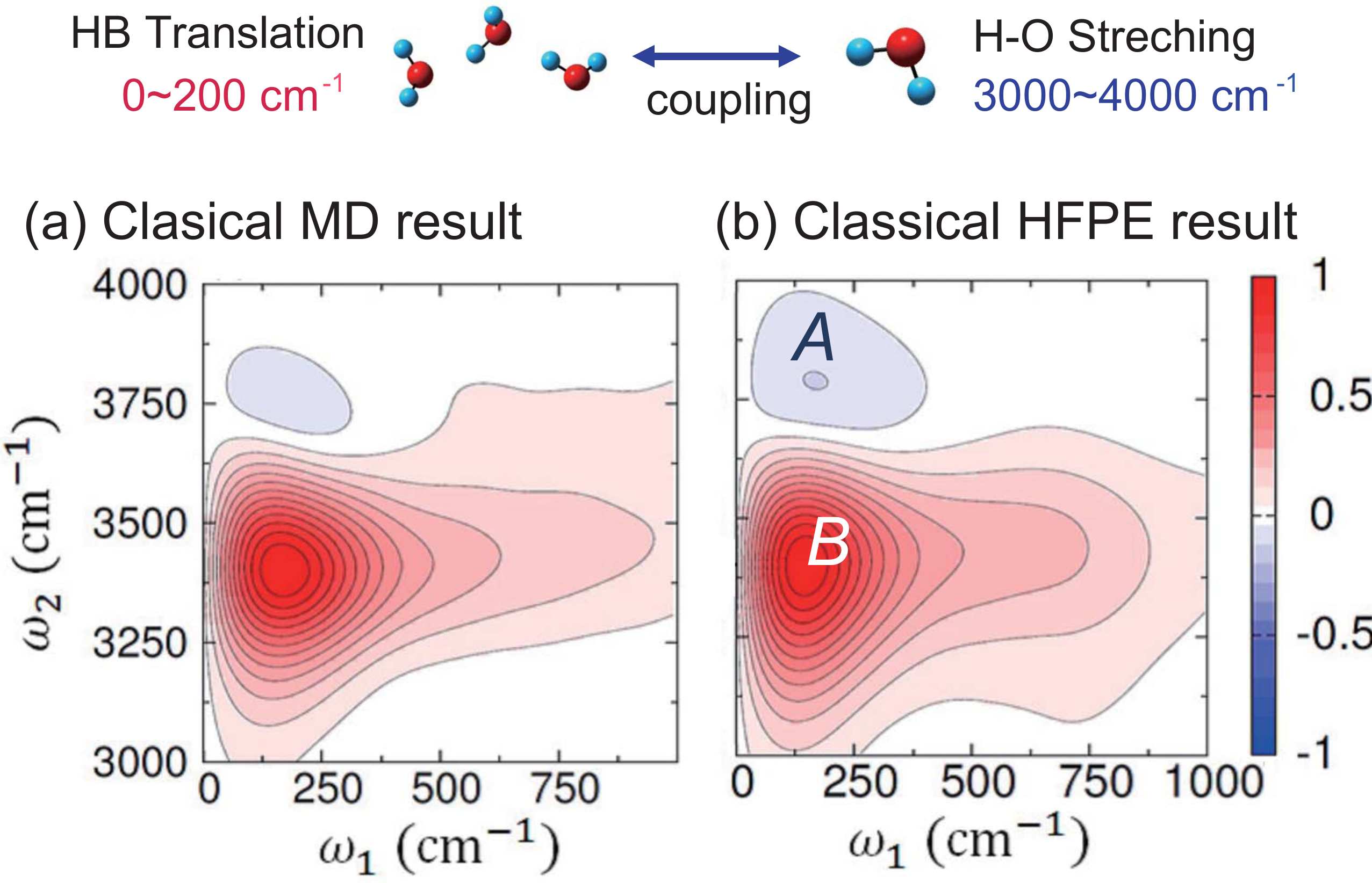}
\caption{\label{IRRaman}2D Raman--IR--IR spectra obtained from molecular dynamics (MD) simulation and from the CHFPE on the basis of the multimode nonlinear Brownian oscillator (BO) model, whose parameters were chosen to reproduce 2D IR--Raman spectra obtained from the MD simulation.  (a)  MD result for the contributions of intermolecular hydrogen-bond (HB) and intramolecular OH stretching mode--mode coupling. (b)  CHFPE result.  From the BO model analysis, we found that the existence of the negative peak $A$ indicates strong coupling between the HB and OH stretching modes, while the peak $B$ arises from nonlinearity of the molecular dipole and polarizability among these two modes. 
Reproduced from H. Ito and Y. Tanimura, \href{https://doi.org/10.1063/1.4941842}{J. Chem. Phys.} \textbf{144}, 074201 (2016), with the permission of AIP Publishing.}
\end{figure}

Although it is only relatively recently that the sensitivity of a nonlinear response function has been investigated as a nonlinear quantum measure,\cite{DijkstraPRL10,DijkstraJPSJ12entangle,DijkstraPTRS2012nonMarkov} the importance of non-Markovian effects as well as that of the  bathentanglement  effects had been realized since the 1980s, when the ultrafast dynamics of molecular motion were first measured by ultrafast nonlinear laser spectroscopy.\cite{Mukamel95} Thus, nonlinear spectra, in particular multidimensional spectra, can be used as a measure of   bathentanglement, which characterizes the difference between $\hat \rho_\mathrm{tot}(t)$ and $\hat \rho_A (t) \otimes \hat \rho_B^\mathrm{eq}$. 

The theoretical foundation for the computation of nonlinear spectra is briefly explained in the Appendix. Because the HEOM approach was originally developed to simulate nonlinear spectra,\cite{Tanimura89B,Tanimura89C, TanimuraMaruyama97, MaruyamaTanimura98} it is able to calculate  nonlinear response functions, which the conventional reduced equation of motion approaches cannot do, as illustrated in Figs.~\ref{fig:nonMarkov}(d) and \ref{2DES}

Since the 1990s, 2D Raman and 2D terahertz Raman vibrational spectra for intermolecular modes\cite{TaniIshiACR09,SteffenTanimura00,TanimuraSteffen00,KatoTanimura02,KatoTanimura04,Ikeda2015} and 2D IR spectra for high-frequency intramolecular modes,\cite{SakuraiJPC11} have been calculated using the QHFPE with  LL + SL system--bath interactions. As shown in Fig.~\ref{IRRaman}, the MD result of 2D IR--Raman was consistently described using the 
classical hierarchical Fokker--Planck  equations (CHFPE) approach and the BO model with anharmonic and nonlinear system--bath coupling.\cite{Ito2016}  An example of a calculated 2D vibrational spectrum (2DVS) is also presented in Fig.~\ref{figPT2DIR:g}.
Because the HEOM formalism treats the quantum and classical systems with any form of potential from the same point of view, it allows identification of purely quantum mechanical effects through comparison of classical and quantum results in the Wigner distribution.\cite{SakuraiJPC11}

The HEOM have also been used to calculate nonlinear electronic spectra. 
By choosing the Liouville paths presented in Appendix for targeting measurements,  the signals of 2D electronic spectra (2DES),\cite{TanimruaJCP12,KramerFMO2DLorentz,Kramer12NJP2Decho,KramerJPC2013BO2DEcho, Yan2011_ChineseJ,Yan2010_2DES,Yan2013_2DES,Shi11FMO2DES,Dijkstra2015fifth,DijkstraNJP10DNA,Dijkstra2019_2DBOS,DijkstraProkorenko2017} were calculated using the regular HEOM, while 2D electronic and vibrational (2DEVS) spectra were calculated using the MS-QHFPE\cite{TanimuraMaruyama97, MaruyamaTanimura98,Ikeda2017JCP} and LT-MS-QFPE. \cite{Ikeda2019JCP, Ikeda2019Ohmic} Examples of calculated nonlinear spectra for pump--probe spectroscopy and transient absorption spectroscopy are presented in Figs.~\ref{fig6MoterPPTA}(b) and \ref{fig6MoterPPTA}(c).

At present, nonlinear spectroscopic measurements are the only experimental approach for observing the ultrafast dynamics of molecular motions, including exciton and electron transfer, and so it is important to calculate nonlinear spectra to confirm the descriptions provided by models.\cite{Dijkstra2015fifth}

Because the HEOM approach does not have to employ the eigenstate representation of the system, it can treat any profile of time-dependent external fields, for which the eigenstate of the system cannot be defined. Thus, optical Stark spectroscopy,\cite{TanimuraMukamelJPSJ94,TanimuraMaruyama97,MaruyamaTanimura98}
double-slit experiments,\cite{Gelin2031doubleslit} and optimal control problems\cite{Desouter-Lecomte2018OptimalControll} have been investigated simply by integrating the HEOM with  time-dependent perturbations.

As an extension of the stochastic approach, the HEOM has been used to calculate linear and nonlinear NMR spectra\cite{TanimuraJPSJ06, Tsuchimoto2015,Joutsuka2008} and muon spin  ($\mu$SR) spectra.\cite{TanimuraJPSJ06, TakahashiJPSJ2020}

\begin{figure}[t]
\includegraphics[width=\columnwidth]{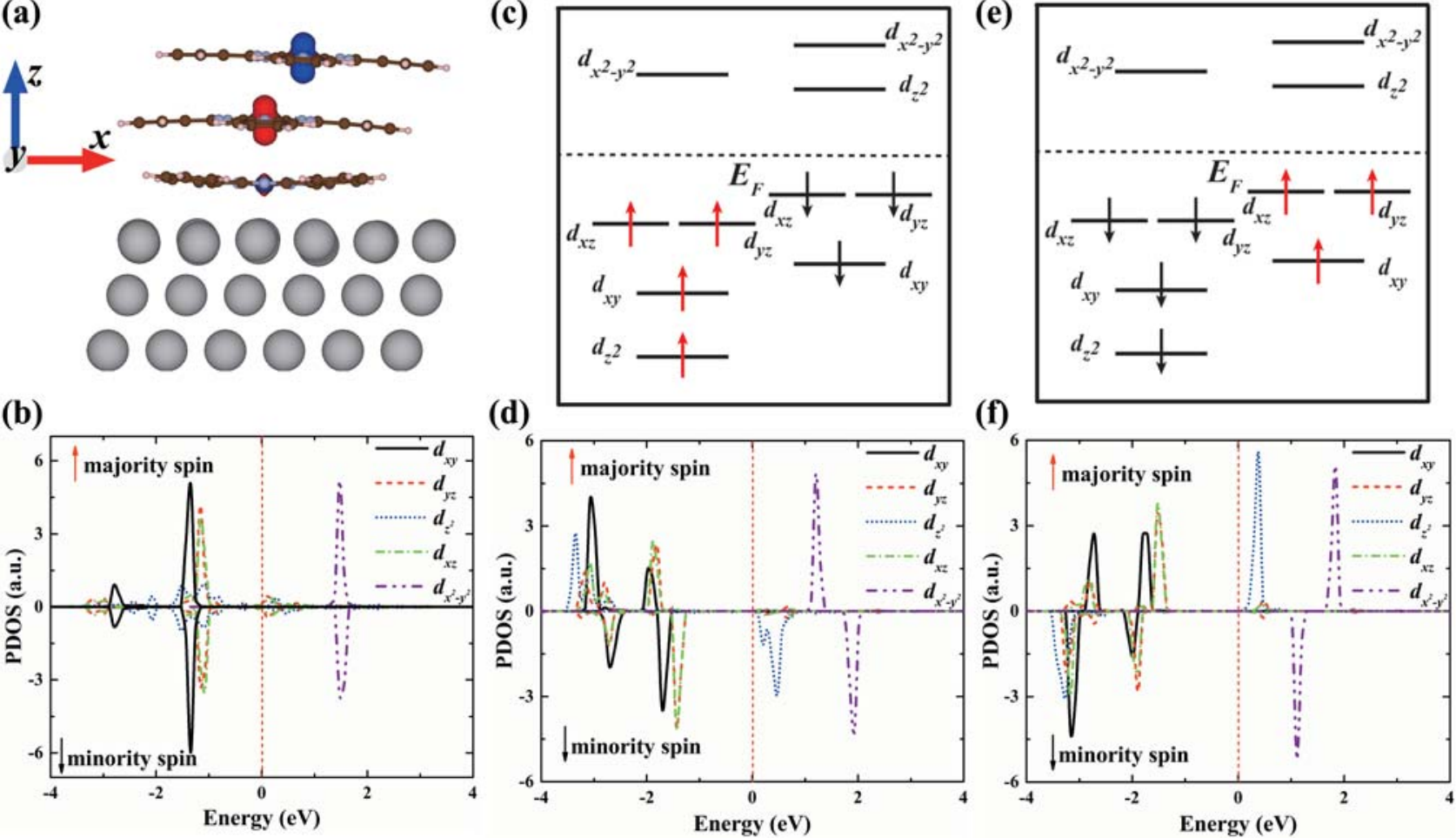}
\caption{(a) Side view of the optimized structure of the triple-layer CoPc/Pb(111) composite calculated from DFT-HEOM. The spin density distribution is also shown, with the isosurface for a spin density of 0.01 ($-0.01$)~\AA$^{-1}$ shaded in red (blue). The lower panels display the PDOS of Co $d$ orbitals in the (b) first, (d) second, and (f) third layers. (c) and (e) Schematic diagrams of the Co $3d$ orbitals in the second and third layers, respectively.
Reproduced from Y. Wang, X. Zheng, and J. Yang, \href{https://doi.org/10.1063/1.4964675} {J. Chem. Phys.} {\bf 145}, 154301 (2016), with the permission of AIP Publishing.
} \label{DFTHEOM}
\end{figure}

\subsection{Quantum transport problems}
At present, there are two HEOM approaches for the treatment of quantum transport problems. One is based on the Wigner distribution function, where the particles can come in or come out through the boundary condition of the system. The heat bath then acts as the thermal energy source for the system. The other is based on the grand canonical treatment of the heat baths, in which the baths act as infinite sources or absorbers of particles and the flow of the particles among the baths is controlled by the chemical potentials. 

The Wigner-function-based approach has been applied to the study of chemical reactions,\cite{TanimuraJCP92} quantum ratchets,\cite{KatoJPCB13} molecular motors,\cite{Ikeda2019JCP} (see Fig.~\ref{fig6MoterPPTA}), photodissociation,\cite{TanimuraMaruyama97,MaruyamaTanimura98}  and resonant tunneling processes\cite{SakuraiJPSJ13,SakuraiNJP14, GrossmannJPC14} (see Fig.~\ref{RTD:g}). Calculations have been performed using the periodic, open, and  inflow--outflow conditions. The investigation of  quantum ratchets showed that the current is suppressed as the tunneling rate increases.\cite{KatoJPCB13} In the investigation of the resonant tunneling diode, a self-current oscillation was discovered in the negative-resistance region, although the Hamiltonian is time-independent (see Fig.~\ref{RTD:g}).

The approach based on the grand canonical ensemble has been applied to inelastic quantum transport problems modeled by a molecular system coupled to  electron baths.\cite{JinYanFermi08, Yan2009YTelecton,Yan2012impurity, Yan2016e-vApprox, StochasticFermi2020A, StochasticFermi2020B} More specifically, Coulomb blockade.\cite{Yan08ClombBlock} the Kondo effect,\cite{ZhengYanFermi08,YanKondo09,Yan2015Kondo,Yan2015EPL} quantum dots,\cite{Reichman2013qdot,Hartle2014} Aharonov--Bohm interferometers, \cite{YanOptimalBasis2015, Yan2015ABeffectB} Anderson localization,\cite{Reichman2015Andersonimp,Okamoto2016Andersonlocal}  
molecular junctions,\cite{Thossperspective2018,Thoss2018transport} and vibronic reaction dynamics at metal surfaces\cite{Thoss2019surfacet} have been studied.
By combining  density functional theory with the HEOM  (DFT-HEOM), the electrical structure of graphene nanoribbons\cite{DFTHEOMChen2013} and the local spin states of adsorbed and embedded organometallic molecules that exhibit Kondo phenomena [including d-CoPc/Au(111),\cite{DFTHEOMJYANG2014} FePc/Au(111),\cite{DFTHEOMJYANG2016} few-layer CoPc/Au(111), \cite{DFTHEOMYANG2016} Au--Co(tpy-SH)$_2$--Au composites, \cite{1stHEOMYAN2016,DFTHEOMYANG2019} CoPc and FePc on Pb(111), \cite{DFTHEOMYANG2018PCCP} and FeOEP on Pb(111)\cite{DFTHEOMYAN2018}] have  been investigated (see Fig.~\ref{DFTHEOM}). In this treatment, the system can dissipate or obtain  energy only through  exchange of  particles with the baths.

\subsection{Quantum information and quantum thermodynamics}\label{subsec:quantum_inf}

\begin{figure}[t]
\includegraphics[width=0.7\columnwidth]{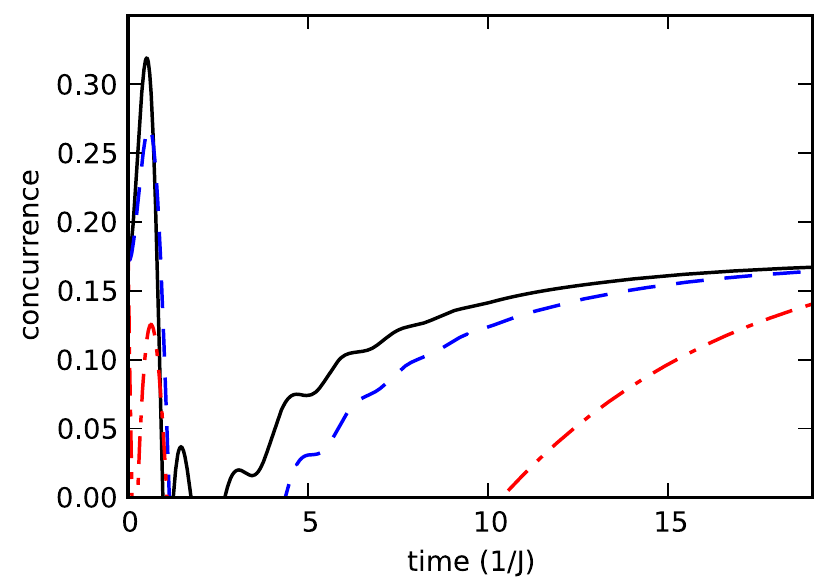}
\caption{Decay of quantum entanglement (concurrence) for two qubits coupled to independent baths as a function of time from a maximum entangled state. We present the results calculated from the HEOM with the  bathentanglement  initial condition (solid curve), the HEOM from factorized initial condition (blue dashed curve), and the Markovian Redfield equation with the RWA (red dash--dotted curve). 
Reproduced with permission from A. G. Dijkstra and Y. Tanimura, \href{https://doi.org/10.1103/PhysRevLett.104.250401}{Phys. Rev. Lett.} {\bf 104}, 250401 (2010).
Copyright 2010 The American Physical Society. } \label{concurrence}
\end{figure}

Because the HEOM can provide an exact numerical treatment of the dynamics defined by the Hamiltonian, it is possible to carry out desktop experiments to verify fundamental statements about quantum information and quantum thermodynamics in a practical manner under  extreme quantum conditions. Commonly studied problems of quantum thermodynamics are quantum heat flow\cite{Katoflow2015,Shi2017flow,HsiehCao2019,Duan2020DuanCao, Thoss2020} and quantum heat engines.\cite{Katoengine2016, Katobook2019} Because the quantum systems we consider are on the nanoscale, we cannot ignore the quantum contributions from the system--bath interaction. Thus, for example, it has been shown that the second law of thermodynamics  is broken if we define the heat current using the system energy instead of the bath energy.\cite{Katoflow2015}

As a problem of quantum information,  quantum entanglement is important not only for the system itself but also for the system--bath interaction. A frequently used quantum measure for a two-qubit system is the concurrence. From the HEOM approach, we observe a rapid revival of the concurrence after the sudden death of entanglement, because the quantum entanglement between the two qubits builds up through the  bathentanglement  between the spins and the baths (see Fig.~\ref{concurrence}). The sensitivity of a nonlinear response function has also been investigated as a nonlinear quantum measure. \cite{DijkstraPRL10,DijkstraJPSJ12entangle} The effects of geometrical phase\cite{Soba2020} and quantum synchronization\cite{Zhang2020} have been investigated simply by integrating the HEOM with  time-dependent driving fields. 

\section{Future Perspectives}
\label{sec:Future}
Thirty years have passed since  HEOM theory was developed as a bridge between the perturbative and Markovian quantum master equation theory and phenomenological stochastic theory. With this in mind, we summarize here the limitations and major challenges of the currently available HEOM theories.

The advantage of the HEOM approach lies in the structure of the equations of motion: a set of simultaneous  differential equations for the elements of the hierarchy  can utilize the  bathentanglement  even under a strong time-dependent external force. When the external perturbation is switched off, the system approaches the  correct thermal equilibrium state at finite temperature, which is important for the  study of relaxation dynamics. The HEOM approach allows us to calculate nonlinear response functions, which is significant for the detection of dynamical bathentanglement. Tremendous efforts have been devoted by many researchers to devise extensions of the HEOM approach that will aid in  the development of theoretical background, numerical techniques, and applications, and thereby further the study of open quantum dynamics. 

Thus, we can now study reasonably large systems,  including those consisting of 20--30 spins and those described by an anharmonic two-dimensional potential, in a numerically ``exact'' manner under  quantum mechanically extreme conditions. While the capability of this approach is still limited, it should become possible to utilize such extended degrees of freedom to provide descriptions of  realistic systems  and of  realistic anharmonic heat baths. 

For example, on the system side, we should be able to employ a quantum chemistry approach or a first-principles MD approach to describe the system dynamics, while the effects of the molecular environment are modeled using a harmonic oscillator bath. Machine learning approaches will be helpful for the construction of realistic system--bath models.\cite{Ueno2020}

On the bath side, 
to treat  non-Gaussian heat baths and spin or correlated fermionic baths, a practical approach is to introduce a reasonably large subsystem between the system and the bath that consists of an ensemble of anharmonic oscillators, spins, or electrons that can interact with each other. If the subsystem degrees of freedom are sufficiently large and the interactions between the subsystem and bath are weak, we should be able to study the effects of an anharmonic or non-bosonic bath whose thermal states are not characterized by the fluctuation--dissipation theorem. Using a model of this kind, we can investigate, for example, the quantum dynamics of an impurity in a complex molecular matrix that is in turn coupled to a phonon bath. The numerical cost of facilitating the treatment of large subsystems is  extremely high, however, and therefore efficient algorithms have to be employed.

The key feature of the HEOM formalism arises from the definition of the hierarchical elements, where the 0th hierarchical element includes all orders of system--bath interactions 
and  is the exact solution of the total Hamiltonian, with the higher members then including the lower-order system--bath interactions, in contrast to conventional perturbative approaches. Although the structure of the HEOM becomes complex and may not be truncated in a simple manner, there is no inherent restriction on constructing the HEOM for any system that interacts with an environment. To study quantum coherence among spatially distributed sites comparable to the thermal de Broglie wavelength, it would be important to construct the HEOM for the Hamiltonian in  momentum space, for example for the Fr\"ohlich Hamiltonian, in the way derived for the deformation potential Hamiltonian.\cite{TanimuraPRA90} In this way, we can explore the interplay of quantum coherence in time and space in a uniform manner.

Finally, it should be mentioned that the framework of the HEOM theory can also be applied to the relativistic quantum theory of the electromagnetic field and to other field theories in which the fundamental interactions arise from the exchange of gauge bosons. The nonperturbative nature of the HEOM approach may provide new insight into the problem of irreversibility not only in time but also in space that we experience in our universe.

\begin{acknowledgments}
The author is grateful to all researchers who have contributed to the development of  HEOM theory. Financial support from the Kyoto University Foundation is acknowledged. 
\end{acknowledgments}

\section*{data availability}
Data sharing is not applicable to this article as no new data were created or analyzed in this study.

\appendix*\label{sec:NRF}
\section{Nonlinear response functions}
\label{sec:NRF}
\begin{figure}
\includegraphics[width=0.8\columnwidth]{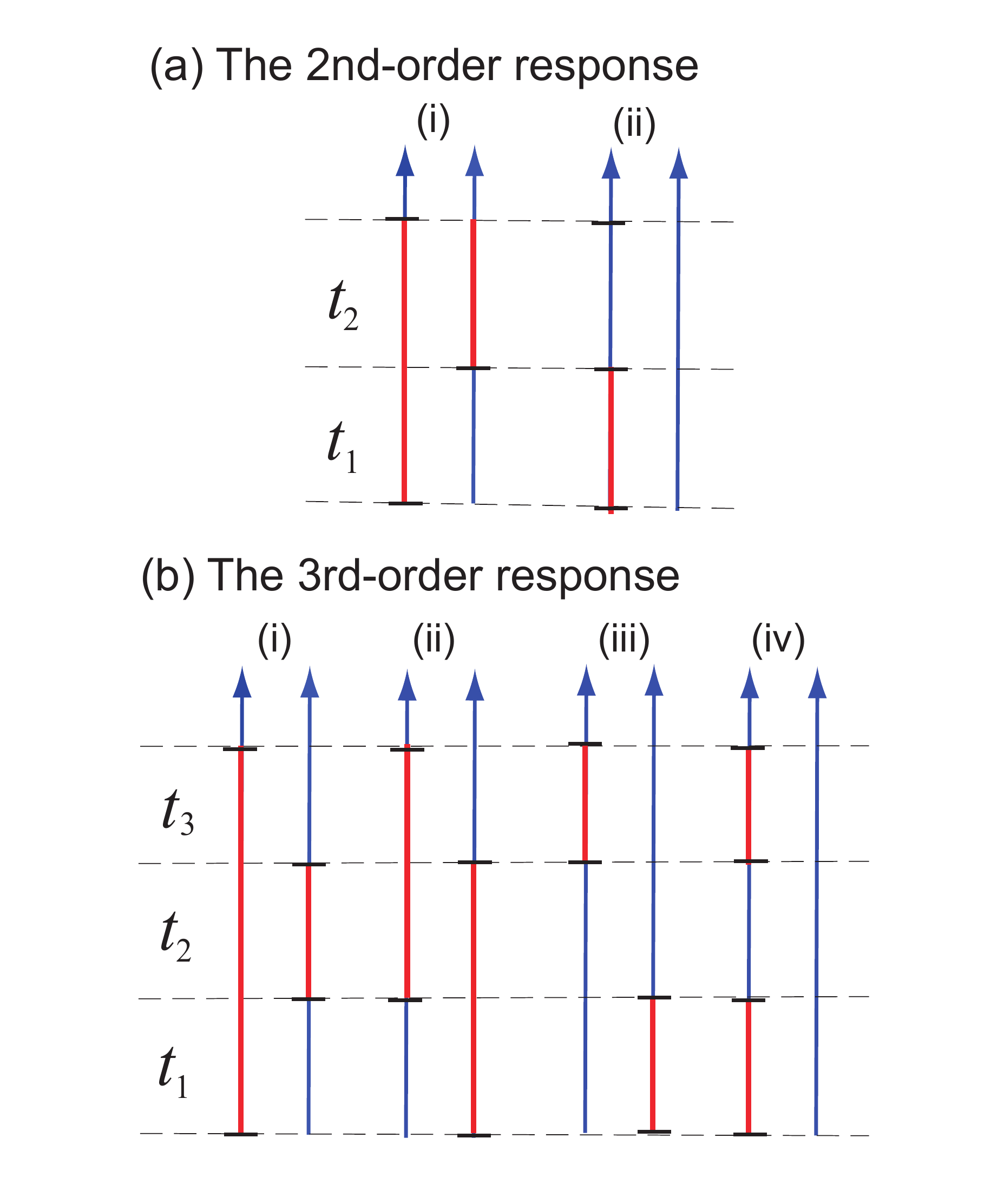}
\caption{\label{LP}Double-sided Feynman diagrams for the second- and third-order response functions: (a) elements of $R^{(2)} (t_2, t_1)$; (b) elements of $R^{(3)} (t_3 ,t_2 ,t_1 )$.  In each diagram,  time runs from  bottom to  top and $t_i$ represents the time intervals for the $i$th sequence between the successive laser--system interactions.  The left line represents the time evolution of the ket, whereas the right line represents that of the bra.  The excited states are represented by the red lines. The complex conjugate paths of these, which can be obtained by interchanging the ket and bra diagrams, are not shown here. (See Ref.~\onlinecite{Mukamel95}.)
\label{Liouvillepath}
}
\end{figure}
\begin{figure}
\includegraphics[width=0.7\columnwidth]{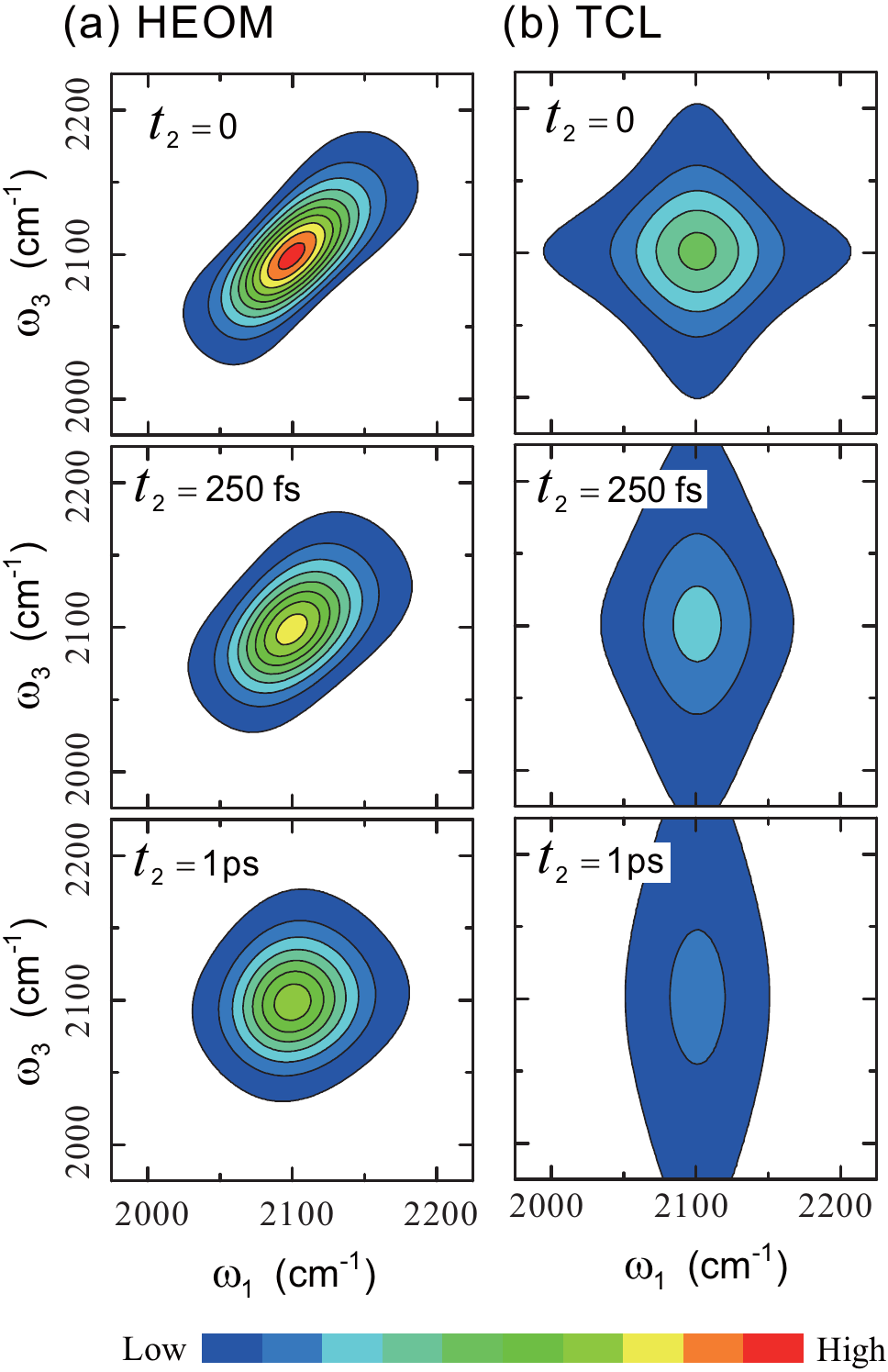}
\caption{\label{2DES}Two-dimensional electronic spectrum of a two-level system calculated from (a) the HEOM approach and (b) the TCL Redfield approach for $\omega_0=2100$~cm$^{-1}$, $\zeta=0.5\omega_0$, $\gamma=0.005\omega_0$, and $\beta \hbar \omega_0 =10$. Because the TCL approach cannot properly take  account of the system--bath coherence (entanglement) over the $t_2$ period, as illustrated by the blue dashed lines in Fig.~\ref{fig:entangle}(b-i), it does not account for the dephasing of the echo signal along the $\omega_1 = \omega_2$ direction. Reproduced with permission from A. Ishizaki and Y. Tanimura, \href{https://doi.org/10.1016/j.chemphys.2007.10.037} {Chem. Phys.} {\bf 347}, 185 (2008). Copyright 2008 
Elsevier.
}
\end{figure}

In quantum mechanics, any physical observable can be expressed as an expectation value of a physical operator. In an optical measurement, the observable at time $t$ is expressed as $P(t)=\mathrm{tr}\{ \hat \mu {\hat \rho}_\mathrm{tot}(t)\}$ or $P(t)=\mathrm{tr}\{ \hat \Pi {\hat \rho}_\mathrm{tot}(t)\}$, where ${\hat \rho}_\mathrm{tot}(t)$ involves the interaction between the excitation fields and the system through the dipole operator $\hat \mu$ or the polarizability operator $\hat \Pi$.\cite{TanimuraJPSJ06} 
We expand $\hat \rho_\mathrm{tot} (t)$ in terms of the $n$th-order response function, which involves  $n$  excitations. In laser spectroscopy, any order of the response function is thus expressed by  $2^{n}$ elements with different configurations corresponding to different time evolutions of the density matrix element.\cite{Mukamel95} The first-order contribution is the linear response function, $R^{(1)}(t_1)= \langle [\hat \mu(t_1), \hat \mu] \rangle/\hbar$. The signal for the harmonic case is presented in Fig.~\ref{fig:nonMarkov}(c). Using  Liouville notation, this is expressed as $R^{(1)} (t_1) = - i\, \mathrm{tr}\{ {\hat \mu \hat G (t_1) \hat \mu ^{\times}  \hat \rho^{eq} } \}/\hbar$, where $\hat G(t)$ is the propagator, which  does not involve the laser interactions.\cite{Tanimura89B}

Each time evolution for the second- and third-order cases is characterized by the quantum Liouville paths for $n=2$ and $3$, as illustrated in Fig.~\ref{Liouvillepath}. The second-order contribution is the observable of  2D Raman, 2D terahertz Raman and 2D IR Raman spectroscopy. For 2D terahertz Raman spectroscopy, the second-order response function is expressed as $R^{(2)}_\mathrm{TTR} (t_2,  t_1) = - \mathrm{tr}\{ {\hat \Pi \hat G (t_2) \hat \mu ^{\times} \hat G (t_1) \hat \mu ^{\times} \hat \rho^{eq} } \} /\hbar^2$.\cite{TanimuraCP98} The diagrams for the second-order response function are presented in Fig.~\ref{Liouvillepath}(a). The complex conjugate diagrams of these are not depicted. The signal for the harmonic case is presented in Fig.~\ref{fig:nonMarkov}(d-i). 

The third-order response function describes 2D infrared (2DIR) spectroscopy as $R^{(3)} (t_3 ,t_2 ,t_1 ) = - i \,\mathrm{tr}\{ \hat \mu  \hat G (t_3 ) \hat \mu ^{\times} \hat G (t_2) \hat \mu ^{\times}  \hat G (t_1) \hat \mu ^{\times}  \hat \rho^{eq} \}/\hbar^3$, where $\hat \mu$ represents the dipole operator. \cite{SteffenTanimura00,TanimuraSteffen00,KatoTanimura02,KatoTanimura04}
The same expression can describe  third-order electronic measurements, which include  two-dimensional electronic spectroscopy (2DES),  by regarding $\hat \mu$ as the electronic transition dipole operator.\cite{TanimruaJCP12} By using the third-order diagrams, the pump--probe spectrum (PPS) and transient absorption spectrum (TAS) are, for example, calculated from the diagrams presented in Figs.~\ref{Liouvillepath}(b-ii) and \ref{Liouvillepath}(b-iii).

In the HEOM approach, the density matrix is replaced by a reduced one, and the Liouvillian in ${G}(t)$ is replaced using Eq.~\eqref{eq:HEOM} or Eq.~\eqref{heom_wig}. We then evaluate the response function, for example, the 2D Raman--IR--IR response function $R_\mathrm{RII}^{(2)} (t_2, t_1)$ by the following steps:
\begin{enumerate}
\item[(i)] The system is initially in the equilibrium state, which is obtained by numerically integrating the HEOM until a steady state is reached. 
\item[(ii)] The system is excited by the first Raman interaction at $t=0$ by applying $\hat{\Pi}^{\times}$ to all of the hierarchical elements to take into account the system--bath entangled states in Figs.~\ref{dsidedFyenman}(a)--(c) that overlap with the diagrams presented in Fig.~\ref{Liouvillepath}(a), as illustrated in Fig.~\ref {fig:entangle}(a-i). 
\item[(iii)] The time evolution of the perturbed elements is then computed by integrating the HEOM for the time period $t_1$. 
\item[(iv)] The second excitation $\hat{\mu}^{\times}$ is then applied, and the perturbed elements are again computed by integrating the HEOM for the time period $t_2$.
\end{enumerate}
The response function $R^{(2)}_\mathrm{RII} (t_2, t_1)$ is calculated from the expectation value of $\hat{\mu}$. If necessary, we calculate $R^{(2)}_\mathrm{RII} [\omega_1, \omega_2]$ by performing a fast Fourier transform for $t_1$ and $t_2$ [Figs.~\ref{fig:nonMarkov}(iv) and \ref{IRRaman}]. The higher-order response functions can be calculated in a similar manner. If we wish to calculate the response function for a specific Liouville path presented in Fig.~\ref{Liouvillepath}, we selectively apply $\hat \mu$ from the right or left instead of $\hat{\mu}^{\times}$ by adjusting the method outlined in steps (i)--(iv) to the targeting Liouville path.\cite{TanimruaJCP12}  In Fig.~\ref{2DES}, we present the 2DES, which is obtained from the double Fourier transform of $R^{(3)} (t_3 ,t_2 ,t_1 )$ for $t_1$ and $t_3$ for the diagrams in Figs.~\ref{Liouvillepath} (b-i) and \ref{Liouvillepath}(b-iv) calculated from the HEOM approach and the TCL Redfield approach. In the high-temperature case, we can also evaluate the nonlinear response function from the HEOM approach analytically as the products of the resolvent ${G}[s]$, which is the Laplace transform of ${G}(t)$. \cite{Tanimura89B,Tanimura89C,IshizakiJCP05}

The sensitivity of a nonlinear response function has been investigated as a nonlinear quantum measure. \cite{DijkstraPRL10,DijkstraJPSJ12entangle, DijkstraPTRS2012nonMarkov}

\end{document}